\documentclass[reprint,nofootinbib, amsmath,amssymb, aps,]{revtex4-2}
\usepackage{graphicx}
\usepackage{xcolor}
\usepackage{dcolumn}
\usepackage{bm}
 \usepackage{rotating} 

\usepackage{hyperref}

\begin{document}
\newcommand{\meff}{m_{\mathrm{eff}}}
\newcommand{\kappaex}{\kappa_{\mathrm{ex}}}
\newcommand{\avdw}{\alpha_{\mathrm{vdW}}}
\newcommand{\tauth}{\tau_{\mathrm{th}}}
\newcommand{\ncav}{n_{\mathrm{cav}}}
\newcommand{\alphaabs}{\alpha_{\mathrm{abs}}}
\newcommand{\epssf}{\varepsilon_{\mathrm{sf}}}
\newcommand{\rmd}{\mathrm{d}}

\title[Strong entropic interactions between light and motion]{Engineered entropic forces allow ultrastrong dynamical backaction}

\author{Andreas Sawadsky}
\author{Raymond A. Harrison}
\author{Glen I. Harris}
\author{Walter W. Wasserman}
\author{Yasmine L. Sfendla}
\author{Warwick P. Bowen}
\email{w.bowen@uq.edu.au}
\author{Christopher G. Baker}
\affiliation{ARC Centre of Excellence for Engineered Quantum Systems, School of Mathematics and Physics, University of Queensland, St Lucia, QLD 4072, Australia.}%

\date{\today}

\begin{abstract}
When confined within an optical cavity light can exert strong radiation pressure forces. Combined with dynamical backaction, this enables important processes such as laser cooling, and applications ranging from precision sensors to quantum memories and interfaces. However, the magnitude of radiation pressure forces is constrained by the energy mismatch between photons and phonons. Here, we overcome this barrier using entropic forces arising from the absorption of light. We show that entropic forces can exceed the radiation pressure force by eight  orders of magnitude, and demonstrate this using a superfluid helium third-sound resonator. We develop a framework to engineer the dynamical backaction from entropic forces, applying it to achieve phonon lasing with a threshold three orders of magnitude lower than previous work. Our results present a pathway to exploit entropic forces in quantum devices, and to study nonlinear fluid phenomena such as turbulence and solitons.
\end{abstract}
\maketitle


\section*{Introduction}

Light generally interacts only weakly with mechanical objects. However, it has been shown that strong interactions can be achieved by confining the light in an optical cavity and employing a low-dissipation mechanical resonance~\cite{kippenberg_analysis_2005}. These cavity optomechanical devices have enabled new physics such as laser amplification and cooling of mechanical motion~\cite{chan_laser_2011}, and new technologies such as precision sensors~\cite{krause2012high, basiri2019precision, purdy2017quantum, forstner2014ultrasensitive}. Remarkably, it has even been possible to reach quantum regimes, where the interaction can create non-classical states of light and of mechanical motion~\cite{safavi2013squeezed, riedinger2018remote} and can enable quantum memories and interfaces, among other quantum technologies~\cite{delaney2022superconducting, bowen_quantum_2015}.

\begin{figure*}
    \centering
    \includegraphics[width=1.0 \textwidth]{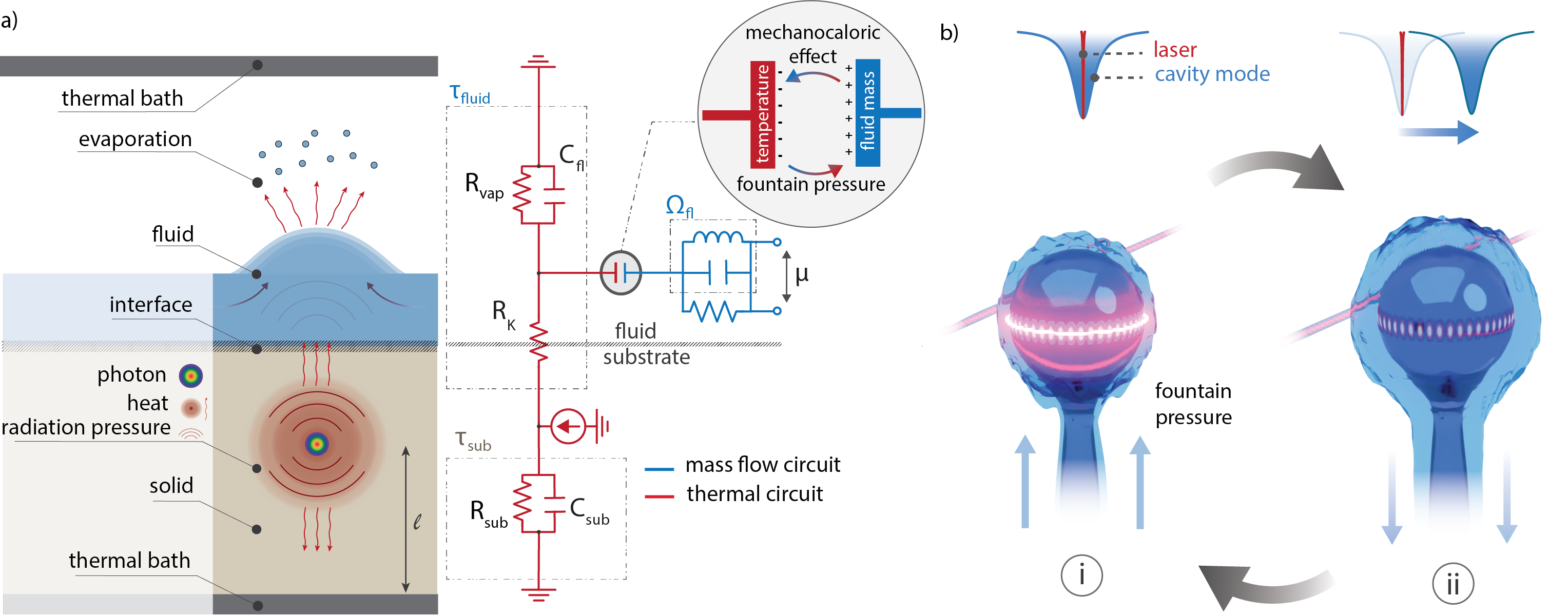}
    \caption{(a) Left: Cross-section schematic showing the superfluid  and heat flow in response to photon absorption. Right: Thermal equivalent circuit (red) coupled to the superfluid  mass flow circuit (blue). R: thermal resistance; C: heat capacity; $\mu$: chemical potential; `K', `sub', `vap', `fl' refer respectively to Kapitza, substrate, vapor and fluid (see Supplementary Material). (b) Superfluid optomechanical resonator used in the experiments: fiber-coupled silica microsphere on a silica stem, covered by a $\sim$24\,nm superfluid film (blue). Illustration of the optomechanical feedback loop between superfluid motion and intracavity power responsible for dynamical backaction~\cite{aspelmeyer_cavity_2014}.}
    \label{fig:thermal_schematic}
\end{figure*}

Usually, the optical forces are achieved using radiation pressure or electrostriction~\cite{kippenberg_analysis_2005,chan_laser_2011}. In these cases, it is generally considered desirable to operate in the {\it resolved sideband regime}, for which the cavity lifetime is longer than the period of the mechanical resonance~\cite{bowen_quantum_2015, aspelmeyer_cavity_2014}. However, 
this prevents multiple scattering processes so that each incident photon can only add or subtract a maximum of one phonon~\cite{aspelmeyer_cavity_2014}. This introduces a fundamental barrier to the strength of forces that can be imparted by light---the maximum efficiency with which energy may be converted from light into mechanical motion is given by the ratio of mechanical to optical frequencies. This ratio is typically in the range of $10^{-9}$ to $10^{-5}$. Multiple scattering is possible in the reverse, {\it non-resolved sideband regime}~\cite{bowen_quantum_2015,aspelmeyer_cavity_2014}. However, in this regime the interaction is suppressed due to the mismatch between the timescales of cavity decay and mechanical motion~\cite{bagheri_dynamic_2011}.

Optical forces can alternatively be applied via entropy gradients, instead of energy gradients. Entropic forces are common in nature, for example, explaining many molecular forces~\cite{asakura_1954, stein_2015} and the restoring force of a stretched rubber band~\cite{treloar1975physics}. Entropic optical forces arise due to absorption of photons, and the resulting heat transfer. Generally known as {\it photothermal forces}, they can allow dynamical backaction effects such as laser cooling of mechanical motion~\cite{metzger_cavity_2004}, including into the quantum regime~\cite{restrepo_classical_2011,de_liberato_quantum_2011}. Interestingly, they can evade the compromise between the number and timescale of interactions that is present for radiation pressure. Operating in the non-resolved sideband regime allows multi-phonon scattering, while the timescale of the interaction can be controlled via the delay introduced during heat transfer~\cite{metzger_cavity_2004,metzger_optical_2008,metzger_self-induced_2008,primo_accurate_2021}. This promises  orders of magnitude stronger optical forces, but has yet to be achieved. This is, in part, due to the complexities of calculating the strength of the interaction and designing devices to leverage it~\cite{primo_accurate_2021}.

Here we demonstrate entropic forces in an optomechanical device that are eight orders-of- magnitude stronger than would be possible with radiation pressure alone. We use them to demonstrate optomechanical phonon lasing with a threshold power of only a few picowatts, a factor of 2000 lower than has been shown before~\cite{meenehan_silicon_2014}. This phonon laser can be viewed as a microscale thermodynamic heat engine. We show that its efficiency---while low due to the small temperature differential introduced by photon absorption---is nonetheless around a hundred times higher than previous nanomechanical heat engines~\cite{wen_coherent_2020, urgell_cooling_2020}, and could be increased above 10\% in purpose-designed devices. Crucial to enabling these results, we develop a new broadly applicable methodology to model optomechanical entropic forces and the dynamical backaction they cause.

Our optomechanical device exploits third sound resonances of a superfluid helium thin film as the mechanical element~\cite{atkins_third_1959,ellis_quantum_1993,harris_laser_2016,sachkou_coherent_2019}. This choice of mechanical element allows the precise engineering of entropic forces, leveraging both entropy-driven fountain pressure, which exists only in superfluids, and the ability to tune the thermal properties of the system over many orders of magnitude via small changes in temperature, film-thickness, and geometry. Combined with the much higher compliance of liquids compared to solids~\cite{he_strong_2020,douvidzon_toward_2021, lee_radiation_2020}, this offers the potential for a broad range of applications, beyond those enabled by dynamic backaction. For instance, the generation of quantized vortices and solitons, as well as all-optical wavelength tuning of microcavities and single photon counting.

\section*{Results}
 
 \subsection*{Modelling and optimising entropic forces} 

Photon absorption events cause a local change in the superfluid temperature, which introduces a fountain (entropic) pressure between hot and cold regions. This drives superfluid flow towards the heat source~\cite{mcauslan_microphotonic_2016, london_thermodynamics_1939,enss_low-temperature_2005}~(the thermomechanical effect~\cite{allen_new_1938}), and in bulk superfluid helium drives normal fluid counterflow. These flows exert forces on the superfluid that can be used to drive superfluid sound modes~\cite{harris_laser_2016}.
Similar to other thermomechanical forces~\cite{metzger_cavity_2004}, the forces are not instantaneous. Rather, they react over the thermal response time $\tau$, which depends on both the properties of the superfluid and the vessel in which it resides. The time delay introduces dynamical backaction that can be used to both cool and amplify the motion of the superfluid~\cite{harris_laser_2016,he_strong_2020}. Optimisation of this entropic backaction requires both that the fountain pressure  is maximized and that

the condition $\Omega \tau \sim 1$ is met, where $\Omega$ is the resonance frequency of the superfluid mode. This condition corresponds to the optimal delay in the optical force for efficient energy transfer~\cite{metzger_self-induced_2008,restrepo_classical_2011}. The fountain pressure is given by $\Delta p = \rho S \Delta T$, where $\rho$, $\Delta T$, and $S$ are respectively the superfluid density, local temperature change, and entropy~\cite{london_thermodynamics_1939}. Employing this relation and extending known results for other thermomechanical forces~\cite{metzger_optical_2008,restrepo_classical_2011,de_liberato_quantum_2011}, we find that the effective dynamical backaction force is given by $F=\rho S \Delta T \mathcal{A} \times\left(\frac{\Omega \tau}{1 + \Omega^2 \tau^2}\right)$, where $\mathcal{A}$ is the effective area of the sound mode, including any spatial mismatch between the mode and the fountain pressure force.

\begin{figure*}
    \centering
    \includegraphics[width=0.95 \textwidth]{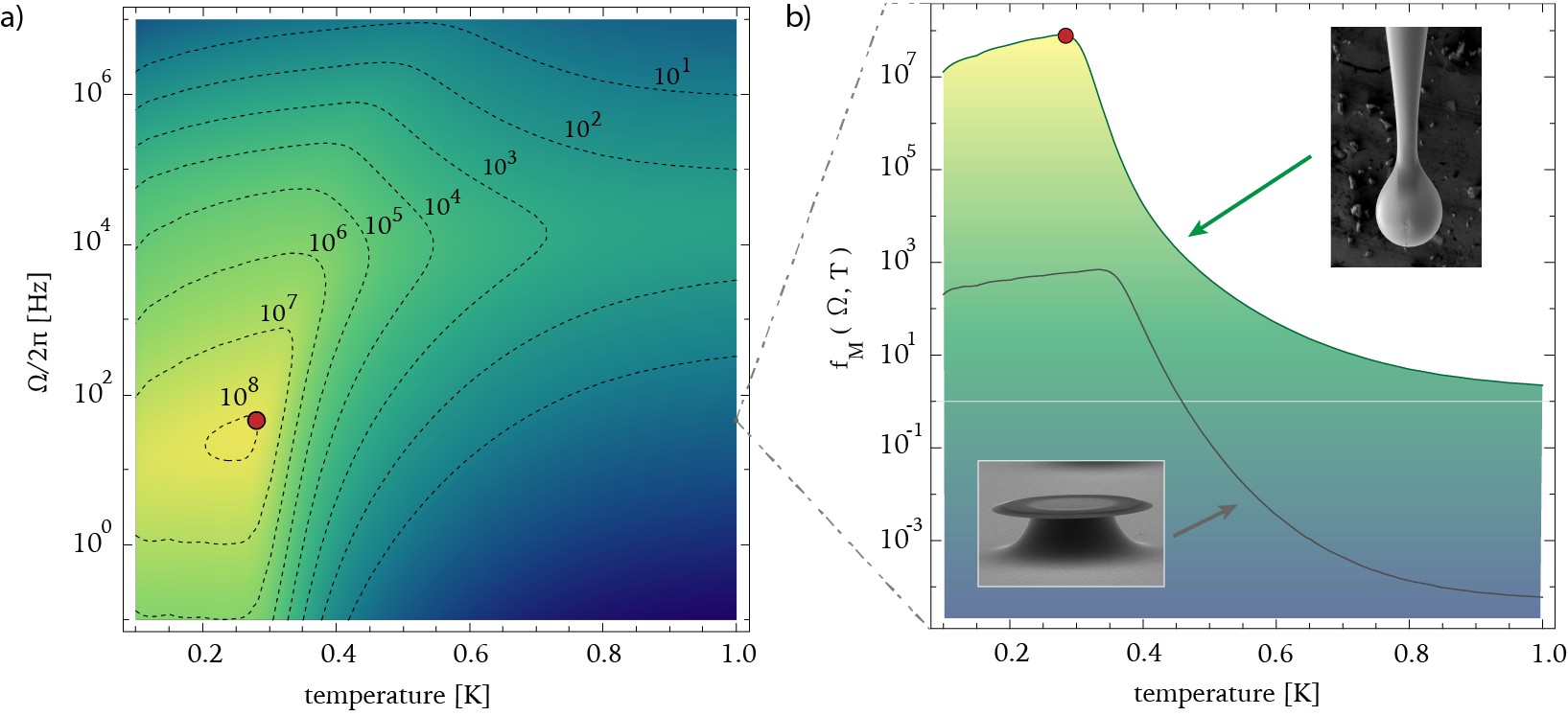}
    \caption{(a) Contour plot of the figure of merit $f_\mathrm{M}(\Omega, T)$, indicating the optimal operating point for enhanced photothermal backaction, as a function of cryostat temperature T and third sound frequency $\Omega/2\pi$. (b) Cut through the dashed white line in (a), showing the figure of merit $f_\mathrm{M}(2\pi\times 72, T)$ for our microsphere resonator (solid line). The grey line plots the same figure of merit for our previous silica microdisk experimental geometry~\cite{he_strong_2020} (calculated with $\Omega=2\pi\times 7$ MHz), for which the optomechanical forces should be entirely dominated by radiation pressure at temperatures above 500\,mK. The horizontal line marks the boundary between the fountain pressure and the radiation pressure-dominated regimes (situated above and below respectively).  Red dot in (a) and (b) marks our experimental operating point.} 
    \label{fig:Contour-Plot_entropy}
\end{figure*}

To model the entropic forces and entropic backaction, we consider the general system geometry shown in Fig. \ref{fig:thermal_schematic}(a). This consists of a superfluid reservoir in contact with a solid substrate and a vapor-phase environment. These, in turn, are in contact with a thermal bath at temperature $T$. Light interacts with the system via photon absorption events which, due to the vanishingly small optical absorption of superfluid helium~\cite{weilert_laser_1995}, are assumed to occur only within the substrate. Heat from absorption events can dissipate either directly into the thermal bath, or by propagating into the superfluid through the interfacial Kapitza resistance and then via evaporation of helium into the vapor-phase.

The Kapitza conductance ($R_K^{-1}$) scales with the  cube of temperature, while the effective thermal conductance of evaporation ($R_{\mathrm{vap}}^{-1}$) increases exponentially with temperature (see Supplementary Material). Small temperature changes can also change the superfluid entropy and thermal conductivity of the substrate by orders of magnitude, while the geometry of the superfluid-coated device can greatly affect the temperature change $\Delta T$ induced by optical absorption, the frequency of the sound mode, and the thermal anchoring of the superfluid. As we will see in the following, these strong dependencies  provide a powerful means to either leverage strong fountain pressure forces at low temperatures, or suppress them at high temperatures so that the unitary radiation pressure interaction can be exploited. 

  \begin{figure*}
    \centering
    \includegraphics[width= .95\textwidth]{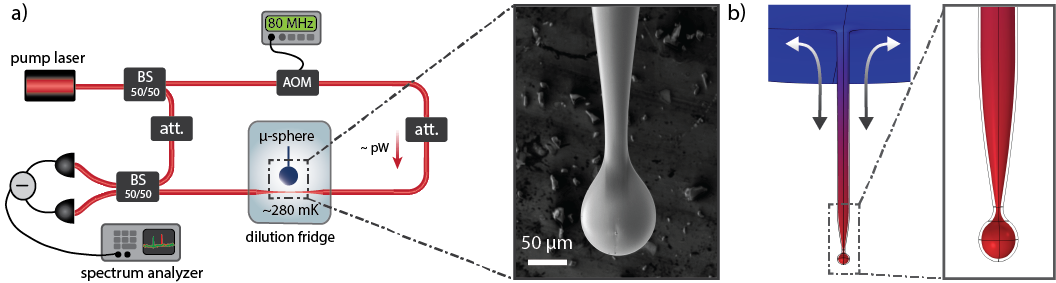}
    \caption{(a) Simplified heterodyne experimental setup schematic, including beam splitters (BS), attenuator (att.) and acousto-optic modulator (AOM). Inset shows a scanning electron microscope (SEM) image of the microsphere. (b) Finite element simulation showing the fundamental superfluid third sound mode confined to the surface of the microsphere resonator (incl. stem), obtained by solving the Helmholtz equation on the resonator's outer surface (see Supplementary Material, section I). Inset shows the displacement profile at the level of the spherical tip, where the WGM field is located.}
    \label{fig:exp_schematic}
\end{figure*}

To enable quantitative predictions, we develop an electric circuit analogue model (right panel, Fig.\,\ref{fig:thermal_schematic}(a)). This model combines a thermal circuit (red) that represents heat flow, a fluid-flow circuit (blue) that represents superfluid mass flow, and a lumped-element RLC circuit that represents the superfluid sound mode. The thermal and fluid-flow circuits are connected via an effective  capacitor which accounts for the fact that a temperature difference drives a superflow via the fountain pressure, while conversely a superflow affects the temperature via the \emph{mechanocaloric} effect~\cite{atkins_evaporation_1959}.  The current, voltage, resistance and capacitance in the electric circuit are respectively the analogs of heat flow $\dot{Q}$, temperature $T$, thermal resistance and heat capacity, while a chemical potential difference $\mu$ drives a mass flow $\dot{m}$ in the superfluid. For a given geometry and temperature, we determine the parameters of the circuit using a combination of existing data and finite-element modelling. We are then able to simulate the dynamics of the sound mode, including entropic forcing and entropic backaction. The simulation ultimately yields a complex transfer function, providing both the amplitude and phase response of the superfluid sound mode (see Supplementary Material, section II). 

As a specific application of our analogue electric circuit model we consider a few nanometer-thick superfluid film self-assembled on a microsphere (Fig.~\ref{fig:thermal_schematic}(b)). Due to the thinness of the film only the superfluid component can move, as the normal fluid component is viscously clamped~\cite{enss_low-temperature_2005}. The superfluid component sustains third sound waves, similar to shallow water waves, due to the van der Waals restoring force between the superfluid and the microsphere surface~\cite{ellis_observation_1989, schechter_observation_1998, harris_laser_2016, baker_theoretical_2016, sachkou_coherent_2019}. The microsphere is suspended on a stem which is attached to a flat substrate. The substrate creates an effective boundary, confining resonant third sound modes. These `stem modes' present as oscillations in the film thickness that flow back and forth along the stem, as illustrated in Fig.~\ref{fig:thermal_schematic}b (i) and (ii). Their resonance frequencies $\Omega$ can be conveniently tuned by changing the length $l$ of the stem, or by tuning the film thickness and therefore the strength of the van der Waals force. Their motion can be optically driven and observed by exploiting optical whispering gallery mode (WGM) resonances that exist within the microsphere~\cite{harris_laser_2016}. 

We choose to study the fundamental stem mode, which features a single antinode at the microsphere and a single node at the substrate. Since the antinode coincides with the position of the optical whispering gallery mode, its motion is well coupled to the WGM resonance. Modelling shows that in our regime the thermal conductivity is such that optical absorption events act to raise the temperature of the entire microsphere-stem system, so that the effective area $\mathcal{A}$ over which the force is applied is, to good approximation, equal to the total surface area of the system. Using these assumptions, we calculate all parameters of the analogue circuit model in Supplementary Material II. This allows us to predict the effective entropic dynamical backaction force $F$.

Our model predicts that there exists an optimal combination of temperature and mechanical resonance frequency to maximise the dynamical backaction force. We find that the optimal resonance frequency coincides with the resonance frequency of the fundamental stem mode if a  stem length of $l \sim 2$~mm is chosen, and that for this stem length the optimal temperature of $T\simeq250$\,mK is easily experimentally accessible. To remove any dependence on laser power, we normalize the dynamical backaction force $F$ by the radiation pressure force $F_{\mathrm{rad}}=\ncav \hbar G$~\cite{aspelmeyer_cavity_2014}, where $\ncav$ is the number of photons within the resonator and $G/2\pi=0.2$ GHz/nm  the radiation pressure optomechanical coupling rate (see Supplementary Material section I). This provides the dimensionless figure of merit $f_M=F/F_{\mathrm{rad}}$ shown in  Fig.\,\ref{fig:Contour-Plot_entropy}(a) as a function of both temperature and mechanical resonance frequency, for a microsphere WGM resonator that we fabricated which has a stem length of $l=1.8$ mm. As can be seen, $f_M$ is predicted to be sharply peaked, varying by more than seven orders of magnitude as the temperature is changed by a factor of ten, and also varying strongly with resonance frequency. At its optimum, the  fountain pressure force is predicted to be over 8 orders of magnitude larger than the radiation pressure force.

To further illustrate the significant gains that can be achieved through optimisation, we calculate the same figure of merit for the microdisk superfluid optomechanical system reported in Ref.~\cite{he_strong_2020}. In that work, phonon-lasing occurred on a high-order Brillouin sound mode at $\Omega/2\pi=7$ MHz. This is shown in Fig.\,\ref{fig:Contour-Plot_entropy}(b) as a function of temperature, alongside the microsphere for comparison. While the trends are similar, the peak value is close to 5 orders of magnitude lower for the microdisk. Our modelling moreover shows that for temperatures in excess of 500 mK, the superfluid becomes much more strongly thermally anchored to the cryostat through its vapor phase than to the substrate. This effectively thermally decouples the superfluid from the microresonator, switching off the fountain pressure interaction. Temperature control therefore provides a useful lever (unexploited in previous work) to either leverage strong fountain pressure forces at low temperatures, or focus on the unitary radiation pressure interaction at higher temperatures. 

\subsection*{Experimental validation} 

To validate the predictions of our model, we locate our microsphere WGM device within a superfluid-tight enclosure whose temperature can be precisely tuned between $T=10$ mK and $T=1.5$ K. A narrow capillary enables the film thickness to be changed in-situ, ranging from a few atomic monolayers to a maximum reached at the saturated vapor pressure, where the sphere is uniformly coated with a $d\simeq 24$\,nm thick superfluid film ($\sim$80 atomic layers), see Supplementary Material I. 
By controlling the bath temperature $T$  and the stem-mode frequency (through superfluid film thickness) we are able to operate at the near-optimal experimental setting where the fountain pressure is maximized and the mechanical frequency matches the thermal response, with $\Omega  \tau = 0.94$, marked by the red dot in Fig. \ref{fig:Contour-Plot_entropy}(a) and Fig. \ref{fig:Contour-Plot_entropy}(b).

\begin{figure*}
    \centering
    \includegraphics[width=0.9 \textwidth]{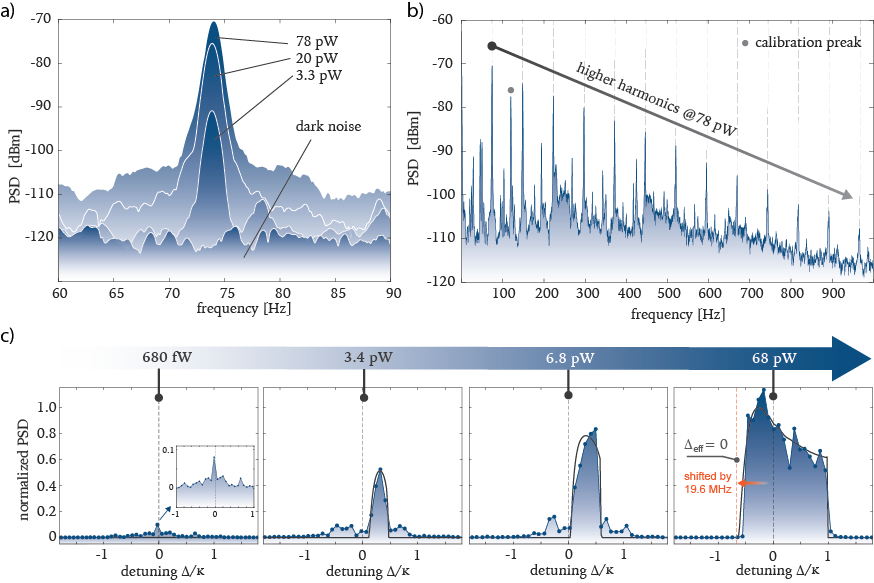}
    \caption{Entropic backaction driven phonon lasing. a) Power spectral density (PSD) of the fundamental stem mode at $\Omega_{\mathrm{M}}/2\pi=72$\,Hz lasing at different input powers. (b) Broader power spectrum at an input power of 78\,pW showing 12 higher harmonics of the  fundamental mode $\Omega_{\mathrm{M}}$. The peak marked by the gray dot is an introduced calibration peak. c) Lasing threshold measurement and the corresponding ODE simulations. Each sub-figure corresponds to different input power, which is increasing from 680\,fW to 68\,pW. Each blue point corresponds to the peak power spectral density at $\Omega_{\mathrm{M}}$ for a  laser detuning $\Delta$ with respect to the cavity resonance, while the black line shows the results of the ODE simulation. At 68\,pW the lasing appears to occur for both positive and negative cavity detunings. This is due to steady-state optical heating of the WGM resonator by around 3.6\,mK above $T=284$\,mK (see  Supplementary Material Fig.\,S10). This DC heating causes an influx of superfluid  shifting the resonance frequency by 19.6\,MHz,  in good agreement with simulations, and corresponds to a tunability of 288 GHz/$\mu$W. }
    \label{fig:lasingthreshold}
\end{figure*}

We optically excite the WGM resonator via evanescent coupling of 1554~nm light from a tapered optical fiber placed in its close proximity, achieving critical coupling to a WGM with intrinsic dissipation rate $\kappa_i/2\pi=15$ MHz. The time-delayed fountain pressure forcing this modifies the dynamics of the fundamental stem mode through the  introduction of an effective optomechanical damping rate $\Gamma_{\mathrm{ent}}$. The sign of this modification may be positive, leading to cooling of the sound mode~\cite{harris_laser_2016}, or negative, leading to amplification~\cite{he_strong_2020}.  In the latter case, phonon lasing occurs once the magnitude of optomechanical (anti-) damping exceeds the intrinsic acoustic damping rate $\Gamma$~\cite{metzger_cavity_2004,metzger_self-induced_2008,restrepo_classical_2011,bowen_quantum_2015}. A measure of the strength of the light-superfluid interaction is provided by the power and intracavity photon number at which this regime is reached.

We employ the heterodyne readout scheme shown in Fig.~\ref{fig:exp_schematic}(a) to measure the power spectra of superfluid oscillation for various input powers. These are shown in Fig.\,\ref{fig:lasingthreshold}(a). The sharp peak observed at the expected frequency of the fundamental stem mode is characteristic of phonon-lasing for cryostat temperatures over the range from  100 mK to 500 mK. We find that the phonon-lasing peak appears for the lowest laser powers at a temperature of around 280 mK, consistent with the predictions of our model. At this temperature the peak is visible even for powers as low as a few picowatts, while for the higher power of 78\,pW more than 10 higher harmonics are seen (see Fig.\,\ref{fig:lasingthreshold}(b)). These harmonics indicate that the phonon lasing has reached motional amplitudes that are sufficiently large to detune the optical cavity significantly away from resonance with the excitation field, thereby introducing a transduction nonlinearity.   

To confirm that the observed behavior is phonon lasing and to constrain the threshold optical drive power, we sweep the laser wavelength over the cavity resonance (from blue- to red-detuned) for various input powers. A characteristic feature of dynamical backaction is that the sign of the optomechanical damping it introduces is opposite on either side of the resonance. We therefore expect to observe lasing on one side of resonance and cooling on the other. For each power and detuning we measure the power spectrum to determine the peak of the dynamic response of the stem mode. In Fig. \ref{fig:lasingthreshold}(c) we plot the normalized peak power spectral density versus detuning for input powers of 680\,fW, 3.4\,pW, 6.8\,pW and 68\,pW.    
At the lowest input power of 680\,fW the measured mode amplitude (blue dots) is low and symmetric around the cavity resonance. This indicates limited dynamical back action. At an input power of 3.4\,pW the measured mode amplitude is about four times higher on the heating side (blue-detuned) compare to the cooling side (red-detuned), indicating that phonon lasing is occurring. At this point, the superfluid oscillator is in a regime of large coherent oscillations with amplitude $\sim$1 \AA, modulating the intensity of the transmitted optical field by more than 60\%. This  provides further confirmation that phonon-lasing is occurring, since, in the absence of other nonlinearities, a large modulation is required to suppress the exponential growth in phonon lasing to a constant amplitude in the steady state. The modulation does this by reducing the average intracavity power, and therefore the steady-state gain of the phonon lasing process~\cite{poot_backaction_2012}. At higher powers again, both the asymmetry and lasing amplitude increase further. These results are in good agreement with numerical simulations (black line) solving the coupled equations of motion for the intracavity field, superfluid motional amplitude and temperature (see Supplementary Material, section III for more details). Our results indicate that the phonon lasing threshold  is in the range of 1 to 3.4 pW, corresponding to an average of less than 0.2 intracavity photons. This threshold is more than three orders of magnitude lower than previous reported values for optically driven optomechanical resonators~\cite{meenehan_silicon_2014}.

Figure~\ref{fig:perspective} compares the phonon-lasing threshold achieved in our work to the state-of-the-art as a function of the effective mass of the mechanical resonator. Prior experiments have used radiation pressure (orange dots), electrostrictive (orange dots), electrothermal (green dots) and photothermal (blue dots)  forces. Most of them form a cluster with microwatt-range threshold powers and picogram-range effective masses. Only electrothermally-driven carbon nanotube resonators reach thresholds approaching a picowatt~\cite{wen_coherent_2020, urgell_cooling_2020}. A significant attribute that enables this is their attogram-level effective mass. This exceptionally low mass allows large motional amplitudes to be driven with small forces. By exploiting engineered entropic backaction, our results operate in a vastly different parameter space, breaking the trend between threshold power and effective mass. Indeed, our 5 gram effective mass is 18 orders of magnitude larger than electrothermally driven systems and 8 orders of magnitude larger than other, much higher threshold, systems. 

\subsection*{Thermodynamic efficiency}
The larger mass of our system compared to other extremely low threshold phonon lasers is partly offset by the superfluid resonator's lower frequency, since the stored mechanical energy scales as $\meff \Omega^2$.  A more complete picture emerges when comparing the systems' thermodynamic efficiencies, which account for the overall optical to acoustical energy conversion.

Our optomechanical phonon laser can be thought of as a heat engine converting heat into mechanical work, with the superfluid film acting as the working medium, the cryostat acting as the cold reservoir and the heated region of the microsphere acting as the hot reservoir. We can estimate its thermodynamic efficiency $\eta$ by comparing the mechanical power $P_{\mathrm{mech}}$ supplied to the superfluid oscillator to sustain its oscillations to the optical power dissipated in the resonator $P_{\mathrm{abs}}$. At an input power of $P=6.8$\,pW, well above the onset of lasing, this yields
$\eta=\frac{P_{\mathrm{mech}}}{P_{\mathrm{abs}}}=\frac{1}{2}\meff \Omega^2 x^2 \Gamma/ (P \times \alpha_{\mathrm{abs}}) \simeq 10^{-5}$ (see Supplementary Material, section V).

This can be compared to the Carnot limit $\eta_{\mathrm{max}}=1-\frac{T_\mathrm{C}}{T_\mathrm{H}}$ determined by the temperature of the hot and cold reservoirs $T_\mathrm{H}$ and $T_\mathrm{C}$. Here $T_\mathrm{C}=T$ is the cryostat temperature, and $T_\mathrm{H}=T+\Delta T$ the temperature of the heated microresonator. At $P=6.8$\,pW, our numerical simulations show that $T_\mathrm{C}=284.0$\,mK and $T\mathrm{_H}=284.4$\,mK (see Supplementary Material section III \& Fig.~S9), such that $\eta_{\mathrm{max}}\simeq 10^{-3}$. This means the system operates within two orders of magnitude of the thermodynamic limit. One reason that it does not reach this limit is that the hot and cold reservoirs are thermally linked through the stem, leading to significant thermal losses.

While the efficiency we achieve is low, to our knowledge it is nonetheless the highest optical-to-mechanical conversion efficiency reported in an optomechanical device.  Indeed, the nonlinear radiation pressure scattering interaction between a photon of frequency $\omega$ and a mechanical oscillator of frequency $\Omega$ leads to an energy transfer $\hbar \Omega$ in the resolved sideband regime (where $\Omega>\kappa$~\cite{aspelmeyer_cavity_2014,bowen_quantum_2015}),  providing an upper bound of $\eta\leq\Omega/\omega$. Gigahertz frequency optomechanical crystals~\cite{meenehan_silicon_2014} operate in a regime where this fraction reaches up to $\Omega/\omega\sim 2.5\times 10^{-5}$, however the total efficiency is suppressed by a factor of $G x/\omega \ll 1$ due to the low rate at which photons enter the cavity via the upper motional sideband~\cite{aspelmeyer_cavity_2014}.
Alternatively, when operating in the unresolved sideband regime ($\kappa \gg \Omega$), each photon can generate many phonons---corresponding to the generation of a large number of higher order motional sidebands. This has been employed to excite mechanical oscillators to very large amplitudes, corresponding to occupations upwards of $10^{12}$ phonons~\cite{bagheri_dynamic_2011}, but the efficiency was limited by the lower phonon energy. In both of the above cases, as well as for low threshold microtoroid resonators~\cite{kippenberg_analysis_2005}, efficiencies are in the  $10^{-8}$ to $10^{-7}$  range.

\begin{figure}
    \centering
    \includegraphics[width=0.47 \textwidth]{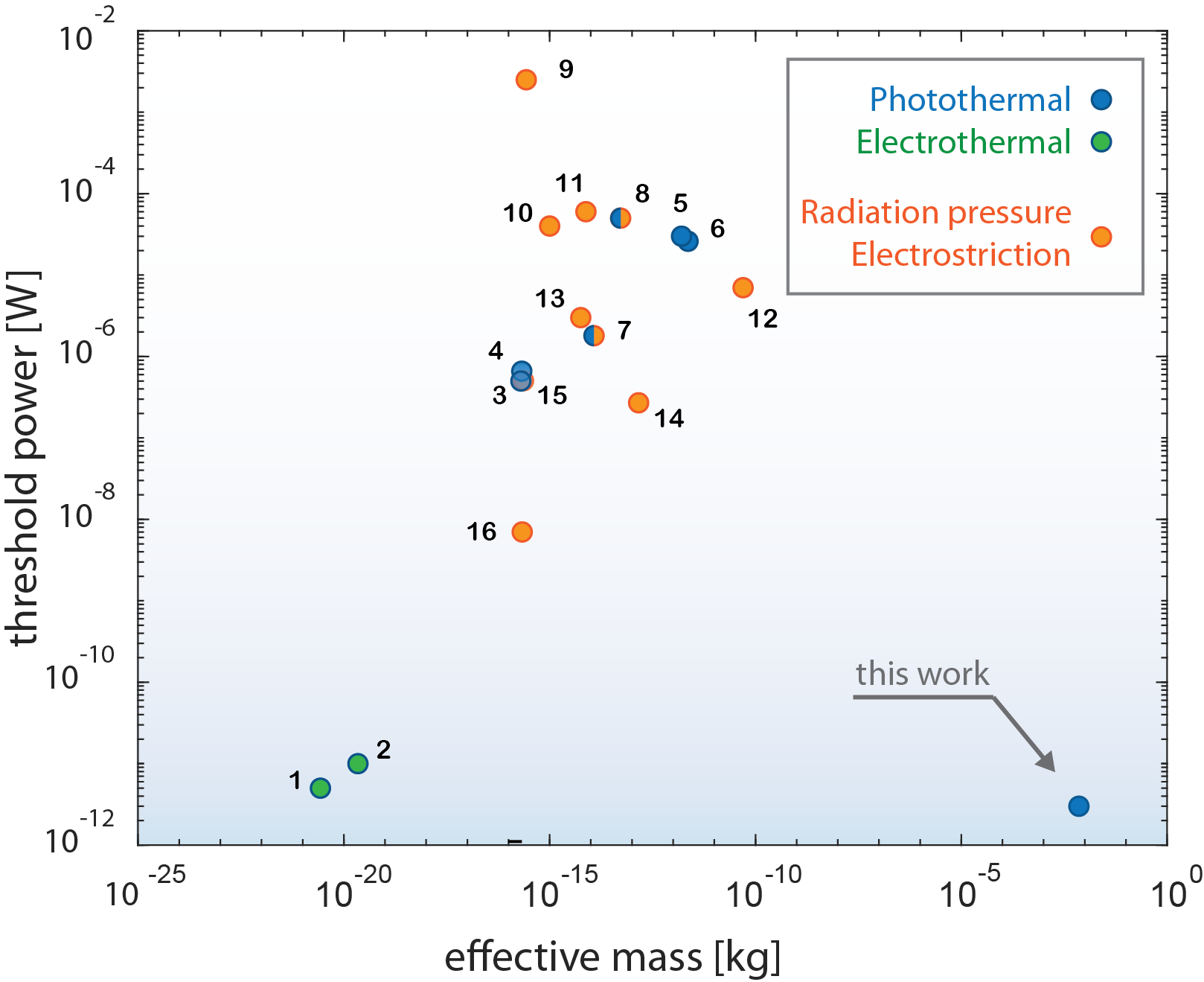}
    \caption{Phonon lasing thresholds of different optomechanical and electromechanical published experiments plotted versus the effective mass of the mechanical mode. The reference publication for each point can be found in Supplementary Material IV.}
    \label{fig:perspective}
\end{figure}

Our efficiency is additionally significantly larger than that  of  photothermally-driven resonators reported in the literature~\cite{metzger_cavity_2004,metzger_self-induced_2008}, where $\eta$ was on the order of $10^{-12}$. It is also two orders of magnitude higher than the efficiency of $10^{-7}$ achieved for an electrothermally-driven carbon nanotube in Ref.~\cite{urgell_cooling_2020}, which has the lowest previously reported phonon-lasing threshold (data point 1 in Fig.\,\ref{fig:perspective}, and Supplementary Material section\,V).

The higher efficiency reported here can be understood by considering that our optomechanical device operates in the unresolved sideband regime, enabling efficient on-resonance  pumping of the cavity, while simultaneously operating in the resolved sideband regime as far as the time delay is concerned.
Indeed, in radiation pressure driven systems, the optimal time-delayed forcing criterion $\Omega \tau_{\mathrm{rp}}\sim 1$ is equivalent to $\Omega\sim \kappa$ (resolved sideband regime), which consequently requires inefficient off-resonant  driving of the cavity.
Here, since the delay is no longer provided by the finite lifetime of the photons in the cavity $\tau_{\mathrm{rp}}=1/\kappa$, but by an independently controllable thermal response time $\tau$, this tension is relieved, enabling a combination of on resonance driving and efficient dynamical backaction  $\Omega \tau \sim 1$.  This allows us to employ a broad optical resonance enabling a  greater dynamic range~\cite{poot_backaction_2012} and larger mechanical amplitudes to be reached, while maintaining a large  time delay $\tau$ for the optical forcing (equivalent to  what would be achieved by an optical cavity with Q of $2\times 10^{12}$).

The thermodynamic efficiency could be further improved by reducing the volume of the resonator. Indeed, the microsphere volume, including the stem, is on the order of $10^6$~$\mu$m$^3$. This can be reduced by approximately six orders of magnitude to $ \sim 1\,\mu$m$^2$ through the use of smaller mode volume resonators such as photonic crystal resonators~\cite{choi_self-similar_2017} or high index WGM cavities~\cite{gil-santos_high-frequency_2015}. The reduced thermal mass then leads to a smaller ratio of $T_C/T_H$, and predicted efficiencies $\eta_{\mathrm{max}}>10$ \%, commensurate with the highest reported values for the efficiency of superfluid fountain-effect pumps~\cite{kittel_orbital_1986}.  

\section*{Discussion}

As discussed earlier, at the operating point of our experiments the entropic forces are around $10^8$ stronger than the radiation pressure forces. Moreover, the dimensionless term $ \left(\frac{\Omega \tau_{\mathrm{rp}}}{1 + \Omega^2 \tau_{\mathrm{rp}}^2}\right)$,  where $\tau_{\mathrm{rp}}=1/\kappa$ is the  photon decay rate,  which quantifies the efficiency of the time-delayed  radiation-pressure forcing for dynamical backaction~\cite{metzger_cavity_2004,metzger_optical_2008,restrepo_classical_2011,de_liberato_quantum_2011, bowen_quantum_2015} is only $2\times 10^{-6}$ for the acoustic mode considered here. In contrast  the equivalent term for fountain pressure is of order 1/2. Combined, this means that the optimized fountain pressure dynamical backaction is over 13 orders of magnitude more effective, enabling the efficient driving of low frequency superfluid acoustic modes in a regime where the radiation pressure fails. We expect this to have  important consequences for the study of superfluid nonlinear hydrodynamical phenomena such as solitonic behaviour~\cite{kurihara1981large} and vortex stirring~\cite{sachkou_coherent_2019}, where the ability to drive and monitor large amplitude wave motion without boiling the film is critical.

It is natural to ask whether the strong forcing achieved here is due to a unique property of the fountain pressure force. We note however that the fountain pressure per absorbed photon is of similar magnitude to the thermoelastic stress which would occur in a crystalline material (see Supplementary Material). It is thus primarily  the engineered ability to collect the thermal energy and operate near the optimal regime $\Omega \tau\sim 1$, combined with the compliance of the fluid interface  which is responsible for the ultralow threshold observed here, and not a fundamentally stronger nature of the fountain pressure force in superfluid helium, making these results broadly applicable to photothermally- and electrothermally-driven systems. The model presented here can also be applied to a variety of different fluids, with some straightforward modifications.

Compared to solids, fluids provide a pathway towards strongly enhanced light-matter interactions, thanks to their compliant interfaces. Early work by Ashkin et al. used radiation pressure from a focussed laser beam to deform an air-liquid interface, using the direction of the deflection (outward) to answer open questions about the momentum of photons in a dielectric~\cite{ashkin_radiation_1973}. The malleability of liquid-liquid and liquid-air interfaces has also been leveraged in optofluidic devices such as adaptive lenses and `lab-on-chip' applications~\cite{monat_integrated_2007}, as well as for fluidic optomechanical resonators~\cite{dahan_droplet_2016}, the generation of giant optical nonlinearities 
~\cite{maksymov_coupling_2019}, and transformable optical cavities~\cite{douvidzon_toward_2021}---which have been proposed to reach the single-photon nonlinear regime \cite{lee_radiation_2020}.
With the thick ($d=24$\,nm) film considered here, the van der Waals pressure $P_{\mathrm{vdW}}=\frac{\rho \alpha_{\mathrm{vdw}}}{d^3}$, which opposes surface deformation and plays the role of the Young's modulus in a solid, is two billion times softer than the underlying silica sphere~\cite{baker_theoretical_2016}. 
The consequence of this compliance extends here beyond dynamical backaction into large static deformations and optical tunability. Indeed, steady-state optical heating of the WGM resonator causes a DC fountain pressure which locally thickens the film. This changes the effective radius of the cavity, shifting the resonance frequency. At an input power of 68\,pW, a DC heating of around 3.6\,mK above the bath temperature of $T=284$\,mK (see Fig.\,S10 in the Supplementary Material) shifts the resonance by 19.6\,MHz,  as shown  in Fig.\,\ref{fig:lasingthreshold} c). This shift corresponds to an exceptionally high all-optical tunability of 288 GHz/$\mu$W, nearly three orders of magnitude larger than that achieved with purpose-built  all optical wavelength routing systems based on mechanically compliant nano-optomechanical resonators \cite{rosenberg_static_2009}.
It is also well in excess of the highest reported values for thermal tuning of microresonators (0.26 GHz/$\mu$W) \cite{sun_45_2016}---albeit with the constraints of cryogenic operation and reduced response rate---underscoring the strength of this optomechanical interaction. In the scenario where the volume of the resonator is reduced to the order of $ \sim 1\,\mu$m$^2$, its thermal mass becomes dominated  by the helium film itself, and the achievable fountain pressure per absorbed photon takes the simple, temperature independent expression $P_{\mathrm{fp}}\simeq\frac{\hbar \omega}{3\,V_{\mathrm{He}}}$. Using this relation, we find that optical frequency shifts larger than $\kappa$ are achievable for a single absorbed photon, potentially providing a novel approach to single photon detection (see Supplementary Material VI).

Finally, this work may lay the groundwork for transformable \emph{phononics}, where optically defined acoustic circuits may be formed on demand. Indeed, DC radiation pressure and/or fountain pressure contributions can substantially thicken the film, as mentioned  in the context of optical tuning. This thickening  also reduces the speed of sound $c_3=\sqrt{3 \avdw d^{-3}}$ within the optically defined `bulge'. Thus, much like an optical waveguide where light is trapped in a higher refractive (lower speed) medium, this could be used to optically trap sound waves~\cite{he_phonon_2018}. Optomechanical backaction in this regime would represent a new form of dynamical
backaction between the intensity of light within a cavity and the shape of the mechanical eigenmodes that it creates, rather than with the momentum of an independently defined mechanical resonator.

\section*{Materials and Methods}

\subsection*{Experimental setup}
The microsphere resonator is formed by melting the tapered end of a silica single-mode fiber (SMF-28) in a fusion splicer. It is then mounted by its non-reflown end inside a superfluid-tight sample chamber within a dilution refrigerator at a temperature $T=284$\,mK.  Infrared laser light ($\lambda=1554$\,nm) from a low-noise erbium-doped fiber laser is evanescently coupled into the microsphere via a tapered optical fiber~\cite{harris_laser_2016}. Helium gas is injected via a capillary to form  a ~$d=24$\,nm-thick saturated superfluid $^4$He film covering the microsphere resonator~\cite{harris_laser_2016,mcauslan_microphotonic_2016,he_strong_2020} (see Supplementary Material, section I for more details). The optical WGMs possess an evanescent component which extends outside the resonator, and are therefore sensitive to the refractive index perturbation caused by the presence and motion of the superfluid~\cite{baker_theoretical_2016}. This leads to a frequency shift of the WGM resonances $\Delta\omega_0=G\times d$, where $G/2\pi=0.2\pm0.01$\,GHz/nm is the optomechanical coupling rate (see Supplementary Material). High sensitivity optical readout of the superfluid's acoustic modes is performed  with a heterodyne detection scheme with a local oscillator field offset by  80\,MHz~\cite{he_strong_2020}, as shown in Fig. \ref{fig:exp_schematic}(a).
Because of the 9 orders of magnitude difference between probe and local oscillator beam powers (pW vs mW), care must be taken to avoid any back-reflections, even as small as those occurring on the facets of the balanced detector. This is done here with a circulator positioned after the microsphere (not shown in Fig. \ref{fig:exp_schematic}(a)).

\section*{Funding}
This work was funded by the US Army Research Office through grant number W911NF17-1-0310 and the Australian Research Council Centre of Excellence for Engineered Quantum Systems (EQUS, project number CE170100009).
C.G.B and G.I.H respectively acknowledge Australian Research Council Fellowships DE190100318 and DE210100848.

\section*{Disclosures}
The authors declare no conflicts of interest.

\section*{Acknowledgements}
The authors acknowledge the facilities, and the scientific and technical assistance, of the Microscopy Australia Facility at the Centre for Microscopy and Microanalysis (CMM), The University of Queensland.


\bibliography{Bibliography}

\begin{thebibliography}{10}

\bibitem{kippenberg_analysis_2005}
T.~Kippenberg, H.~Rokhsari, T.~Carmon, A.~Scherer, and K.~Vahala, ``Analysis of
  radiation-pressure induced mechanical oscillation of an optical
  microcavity,'' {\em Physical Review Letters}, vol.~95, no.~3, p.~033901,
  2005.

\bibitem{chan_laser_2011}
J.~Chan, T.~P.~M. Alegre, A.~H. Safavi-Naeini, J.~T. Hill, A.~Krause,
  S.~Gröblacher, M.~Aspelmeyer, and O.~Painter, ``Laser cooling of a
  nanomechanical oscillator into its quantum ground state,'' {\em Nature},
  vol.~478, pp.~89--92, Oct. 2011.

\bibitem{krause2012high}
A.~G. Krause, M.~Winger, T.~D. Blasius, Q.~Lin, and O.~Painter, ``A
  high-resolution microchip optomechanical accelerometer,'' {\em Nature
  Photonics}, vol.~6, no.~11, pp.~768--772, 2012.

\bibitem{basiri2019precision}
S.~Basiri-Esfahani, A.~Armin, S.~Forstner, and W.~P. Bowen, ``Precision
  ultrasound sensing on a chip,'' {\em Nature Communications}, vol.~10, no.~1,
  pp.~1--9, 2019.

\bibitem{purdy2017quantum}
T.~Purdy, K.~Grutter, K.~Srinivasan, and J.~Taylor, ``Quantum correlations from
  a room-temperature optomechanical cavity,'' {\em Science}, vol.~356,
  no.~6344, pp.~1265--1268, 2017.

\bibitem{forstner2014ultrasensitive}
S.~Forstner, E.~Sheridan, J.~Knittel, C.~L. Humphreys, G.~A. Brawley,
  H.~Rubinsztein-Dunlop, and W.~P. Bowen, ``Ultrasensitive optomechanical
  magnetometry,'' {\em Advanced Materials}, vol.~26, no.~36, pp.~6348--6353,
  2014.

\bibitem{safavi2013squeezed}
A.~H. Safavi-Naeini, S.~Gr{\"o}blacher, J.~T. Hill, J.~Chan, M.~Aspelmeyer, and
  O.~Painter, ``Squeezed light from a silicon micromechanical resonator,'' {\em
  Nature}, vol.~500, no.~7461, pp.~185--189, 2013.

\bibitem{riedinger2018remote}
R.~Riedinger, A.~Wallucks, I.~Marinkovi{\'c}, C.~L{\"o}schnauer, M.~Aspelmeyer,
  S.~Hong, and S.~Gr{\"o}blacher, ``Remote quantum entanglement between two
  micromechanical oscillators,'' {\em Nature}, vol.~556, no.~7702,
  pp.~473--477, 2018.

\bibitem{delaney2022superconducting}
R.~Delaney, M.~Urmey, S.~Mittal, B.~Brubaker, J.~Kindem, P.~Burns, C.~Regal,
  and K.~Lehnert, ``Superconducting-qubit readout via low-backaction
  electro-optic transduction,'' {\em Nature}, vol.~606, no.~7914, pp.~489--493,
  2022.

\bibitem{bowen_quantum_2015}
W.~P. Bowen and G.~J. Milburn, {\em Quantum optomechanics}.
\newblock CRC Press, 2015.

\bibitem{aspelmeyer_cavity_2014}
M.~Aspelmeyer, T.~J. Kippenberg, and F.~Marquardt, ``Cavity optomechanics,''
  {\em Reviews of Modern Physics}, vol.~86, pp.~1391--1452, Dec. 2014.

\bibitem{bagheri_dynamic_2011}
M.~Bagheri, M.~Poot, M.~Li, W.~P.~H. Pernice, and H.~X. Tang, ``Dynamic
  manipulation of nanomechanical resonators in the high-amplitude regime and
  non-volatile mechanical memory operation,'' {\em Nature Nanotechnology},
  vol.~6, pp.~726--732, Oct. 2011.

\bibitem{asakura_1954}
S.~Asakura and F.~Oosawa, ``On interaction between two bodies immersed in a
  solution of macromolecules,'' {\em The Journal of Chemical Physics}, vol.~22,
  no.~7, pp.~1255--1256, 1954.

\bibitem{stein_2015}
X.~Liu, M.~M. Skanata, and D.~Stein, ``Entropic cages for trapping dna near a
  nanopore,'' {\em Nature Communications}, vol.~6, no.~1, p.~6222, 2015.

\bibitem{treloar1975physics}
L.~G. Treloar, {\em The physics of rubber elasticity}.
\newblock Oxford University Press, 1975.

\bibitem{metzger_cavity_2004}
C.~H. Metzger and K.~Karrai, ``Cavity cooling of a microlever,'' {\em Nature},
  vol.~432, no.~7020, pp.~1002--1005, 2004.

\bibitem{restrepo_classical_2011}
J.~Restrepo, J.~Gabelli, C.~Ciuti, and I.~Favero, ``Classical and quantum
  theory of photothermal cavity cooling of a mechanical oscillator,'' {\em
  Comptes Rendus Physique}, vol.~12, pp.~860--870, Dec. 2011.

\bibitem{de_liberato_quantum_2011}
S.~De~Liberato, N.~Lambert, and F.~Nori, ``Quantum noise in photothermal
  cooling,'' {\em Physical Review A}, vol.~83, p.~033809, Mar. 2011.

\bibitem{metzger_optical_2008}
C.~Metzger, I.~Favero, A.~Ortlieb, and K.~Karrai, ``Optical self cooling of a
  deformable {Fabry}-{Perot} cavity in the classical limit,'' {\em Physical
  Review B}, vol.~78, p.~035309, July 2008.

\bibitem{metzger_self-induced_2008}
C.~Metzger, M.~Ludwig, C.~Neuenhahn, A.~Ortlieb, I.~Favero, K.~Karrai, and
  F.~Marquardt, ``Self-{Induced} {Oscillations} in an {Optomechanical} {System}
  {Driven} by {Bolometric} {Backaction},'' {\em Physical Review Letters},
  vol.~101, p.~133903, Sept. 2008.
\newblock Publisher: American Physical Society.

\bibitem{primo_accurate_2021}
A.~G. Primo, C.~M. Kersul, R.~Benevides, N.~C. Carvalho, M.~Ménard, N.~C.
  Frateschi, P.-L. de~Assis, G.~S. Wiederhecker, and T.~P. Mayer~Alegre,
  ``Accurate modeling and characterization of photothermal forces in
  optomechanics,'' {\em APL Photonics}, vol.~6, p.~086101, Aug. 2021.

\bibitem{meenehan_silicon_2014}
S.~M. Meenehan, J.~D. Cohen, S.~Gröblacher, J.~T. Hill, A.~H. Safavi-Naeini,
  M.~Aspelmeyer, and O.~Painter, ``Silicon optomechanical crystal resonator at
  millikelvin temperatures,'' {\em Physical Review A}, vol.~90, p.~011803, July
  2014.

\bibitem{wen_coherent_2020}
Y.~Wen, N.~Ares, F.~J. Schupp, T.~Pei, G.~a.~D. Briggs, and E.~A. Laird, ``A
  coherent nanomechanical oscillator driven by single-electron tunnelling,''
  {\em Nature Physics}, vol.~16, pp.~75--82, Jan. 2020.

\bibitem{urgell_cooling_2020}
C.~Urgell, W.~Yang, S.~L. De~Bonis, C.~Samanta, M.~J. Esplandiu, Q.~Dong,
  Y.~Jin, and A.~Bachtold, ``Cooling and self-oscillation in a nanotube
  electromechanical resonator,'' {\em Nature Physics}, vol.~16, pp.~32--37,
  Jan. 2020.

\bibitem{atkins_third_1959}
K.~R. Atkins, ``Third and {Fourth} {Sound} in {Liquid} {Helium} {II},'' {\em
  Physical Review}, vol.~113, pp.~962--965, Feb. 1959.

\bibitem{ellis_quantum_1993}
F.~M. Ellis and L.~Li, ``Quantum swirling of superfluid helium films,'' {\em
  Physical review letters}, vol.~71, no.~10, p.~1577, 1993.

\bibitem{harris_laser_2016}
G.~I. Harris, D.~L. McAuslan, E.~Sheridan, Y.~Sachkou, C.~Baker, and W.~P.
  Bowen, ``Laser cooling and control of excitations in superfluid helium,''
  {\em Nature Physics}, vol.~12, pp.~788--793, Aug. 2016.

\bibitem{sachkou_coherent_2019}
Y.~P. Sachkou, C.~G. Baker, G.~I. Harris, O.~R. Stockdale, S.~Forstner, M.~T.
  Reeves, X.~He, D.~L. McAuslan, A.~S. Bradley, M.~J. Davis, and W.~P. Bowen,
  ``Coherent vortex dynamics in a strongly interacting superfluid on a silicon
  chip,'' {\em Science}, vol.~366, pp.~1480--1485, Dec. 2019.

\bibitem{he_strong_2020}
X.~He, G.~I. Harris, C.~G. Baker, A.~Sawadsky, Y.~L. Sfendla, Y.~P. Sachkou,
  S.~Forstner, and W.~P. Bowen, ``Strong optical coupling through superfluid
  {Brillouin} lasing,'' {\em Nature Physics}, pp.~1--5, Feb. 2020.

\bibitem{douvidzon_toward_2021}
M.~Douvidzon, S.~Maayani, H.~Nagar, T.~Admon, V.~Shuvayev, L.~Yang, L.~Deych,
  Y.~Roichman, and T.~Carmon, ``Toward transformable photonics: {Reversible}
  deforming soft cavities, controlling their resonance split and directional
  emission,'' {\em APL Photonics}, vol.~6, p.~071304, July 2021.

\bibitem{lee_radiation_2020}
A.~Lee, P.~Zhang, Y.~Xu, and S.~Jung, ``Radiation pressure-induced nonlinearity
  in a micro-droplet,'' {\em Optics Express}, vol.~28, p.~12675, Apr. 2020.

\bibitem{mcauslan_microphotonic_2016}
D.~McAuslan, G.~Harris, C.~Baker, Y.~Sachkou, X.~He, E.~Sheridan, and W.~Bowen,
  ``Microphotonic {Forces} from {Superfluid} {Flow},'' {\em Physical Review X},
  vol.~6, p.~021012, Apr. 2016.

\bibitem{london_thermodynamics_1939}
H.~London, ``Thermodynamics of the thermomechanical effect of liquid {He}
  {II},'' {\em Proceedings of the Royal Society of London. Series A,
  Mathematical and Physical Sciences}, pp.~484--496, 1939.

\bibitem{enss_low-temperature_2005}
C.~Enss and S.~Hunklinger, {\em Low-{Temperature} {Physics}}.
\newblock Springer, Apr. 2005.

\bibitem{allen_new_1938}
J.~F. Allen and H.~Jones, ``New phenomena connected with heat flow in helium
  {II},'' {\em Nature}, vol.~141, no.~3562, pp.~243--244, 1938.

\bibitem{weilert_laser_1995}
M.~A. Weilert, D.~L. Whitaker, H.~J. Maris, and G.~M. Seidel, ``Laser
  levitation of superfluid helium,'' {\em Journal of Low Temperature Physics},
  vol.~98, pp.~17--35, Jan. 1995.

\bibitem{atkins_evaporation_1959}
K.~R. Atkins, B.~Rosenbaum, and H.~Seki, ``Evaporation {Effects} during
  {Superflow} of {Liquid} {Helium} {II},'' {\em Physical Review}, vol.~113,
  pp.~751--754, Feb. 1959.

\bibitem{ellis_observation_1989}
F.~M. Ellis and H.~Luo, ``Observation of the persistent-current splitting of a
  third-sound resonator,'' {\em Physical Review B}, vol.~39, no.~4, p.~2703,
  1989.

\bibitem{schechter_observation_1998}
A.~M.~R. Schechter, R.~W. Simmonds, R.~E. Packard, and J.~C. Davis,
  ``Observation of ‘third sound’ in superfluid {3He},'' {\em Nature},
  vol.~396, pp.~554--557, Dec. 1998.

\bibitem{baker_theoretical_2016}
C.~G. Baker, G.~I. Harris, D.~L. McAuslan, Y.~Sachkou, X.~He, and W.~P. Bowen,
  ``Theoretical framework for thin film superfluid optomechanics: towards the
  quantum regime,'' {\em New Journal of Physics}, vol.~18, p.~123025, Dec.
  2016.

\bibitem{poot_backaction_2012}
M.~Poot, K.~Y. Fong, M.~Bagheri, W.~H.~P. Pernice, and H.~X. Tang, ``Backaction
  limits on self-sustained optomechanical oscillations,'' {\em Physical Review
  A}, vol.~86, p.~053826, Nov. 2012.
\newblock Publisher: American Physical Society.

\bibitem{choi_self-similar_2017}
H.~Choi, M.~Heuck, and D.~Englund, ``Self-{Similar} {Nanocavity} {Design} with
  {Ultrasmall} {Mode} {Volume} for {Single}-{Photon} {Nonlinearities},'' {\em
  Physical Review Letters}, vol.~118, p.~223605, May 2017.

\bibitem{gil-santos_high-frequency_2015}
E.~Gil-Santos, C.~Baker, D.~T. Nguyen, W.~Hease, C.~Gomez, A.~Lemaître,
  S.~Ducci, G.~Leo, and I.~Favero, ``High-frequency nano-optomechanical disk
  resonators in liquids,'' {\em Nature Nanotechnology}, vol.~10, pp.~810--816,
  Sept. 2015.

\bibitem{kittel_orbital_1986}
P.~Kittel, ``Orbital resupply of liquid helium,'' {\em Journal of Spacecraft
  and Rockets}, vol.~23, pp.~391--396, July 1986.

\bibitem{kurihara1981large}
S.~Kurihara, ``Large-amplitude quasi-solitons in superfluid films,'' {\em
  Journal of the Physical Society of Japan}, vol.~50, no.~10, pp.~3262--3267,
  1981.

\bibitem{ashkin_radiation_1973}
A.~Ashkin and J.~M. Dziedzic, ``Radiation {Pressure} on a {Free} {Liquid}
  {Surface},'' {\em Physical Review Letters}, vol.~30, pp.~139--142, Jan. 1973.

\bibitem{monat_integrated_2007}
C.~Monat, P.~Domachuk, and B.~J. Eggleton, ``Integrated optofluidics: {A} new
  river of light,'' {\em Nature Photonics}, vol.~1, pp.~106--114, Feb. 2007.

\bibitem{dahan_droplet_2016}
R.~Dahan, L.~L. Martin, and T.~Carmon, ``Droplet optomechanics,'' {\em Optica},
  vol.~3, p.~175, Feb. 2016.

\bibitem{maksymov_coupling_2019}
I.~S. Maksymov and A.~D. Greentree, ``Coupling light and sound: giant
  nonlinearities from oscillating bubbles and droplets,'' {\em Nanophotonics},
  vol.~8, pp.~367--390, Mar. 2019.

\bibitem{rosenberg_static_2009}
J.~Rosenberg, Q.~Lin, and O.~Painter, ``Static and dynamic wavelength routing
  via the gradient optical force,'' {\em Nature Photonics}, vol.~3,
  pp.~478--483, Aug. 2009.

\bibitem{sun_45_2016}
C.~Sun, M.~Wade, M.~Georgas, S.~Lin, L.~Alloatti, B.~Moss, R.~Kumar, A.~H.
  Atabaki, F.~Pavanello, J.~M. Shainline, J.~S. Orcutt, R.~J. Ram, M.~Popović,
  and V.~Stojanović, ``A 45 nm {CMOS}-{SOI} {Monolithic} {Photonics}
  {Platform} {With} {Bit}-{Statistics}-{Based} {Resonant} {Microring} {Thermal}
  {Tuning},'' {\em IEEE Journal of Solid-State Circuits}, vol.~51,
  pp.~893--907, Apr. 2016.

\bibitem{he_phonon_2018}
X.~He, C.~Baker, Y.~Sachkou, A.~Sawadsky, S.~Forstner, Y.~Sfendla, and
  W.~Bowen, ``Phonon {Confinement} by the {Force} of {Light},'' in {\em {CLEO}
  {Pacific} {Rim} {Conference}}, (Hong Kong), p.~Tu3F.3, OSA, 2018.

\end{thebibliography}


\begin{thebibliography}{10}

\bibitem{he_strong_2020}
X.~He, G.~I. Harris, C.~G. Baker, A.~Sawadsky, Y.~L. Sfendla, Y.~P. Sachkou,
  S.~Forstner, and W.~P. Bowen, ``Strong optical coupling through superfluid
  {Brillouin} lasing,'' {\em Nature Physics}, pp.~1--5, Feb. 2020.

\bibitem{harris_laser_2016}
G.~I. Harris, D.~L. McAuslan, E.~Sheridan, Y.~Sachkou, C.~Baker, and W.~P.
  Bowen, ``Laser cooling and control of excitations in superfluid helium,''
  {\em Nature Physics}, vol.~12, pp.~788--793, Aug. 2016.

\bibitem{ellis_observation_1989}
F.~M. Ellis and H.~Luo, ``Observation of the persistent-current splitting of a
  third-sound resonator,'' {\em Physical Review B}, vol.~39, no.~4, p.~2703,
  1989.

\bibitem{baker_theoretical_2016}
C.~G. Baker, G.~I. Harris, D.~L. McAuslan, Y.~Sachkou, X.~He, and W.~P. Bowen,
  ``Theoretical framework for thin film superfluid optomechanics: towards the
  quantum regime,'' {\em New Journal of Physics}, vol.~18, p.~123025, Dec.
  2016.

\bibitem{matsko_optical_2006}
A.~B. Matsko and V.~S. Ilchenko, ``Optical resonators with whispering-gallery
  modes-part {I}: basics,'' {\em IEEE Journal of Selected Topics in Quantum
  Electronics}, vol.~12, pp.~3--14, Jan. 2006.

\bibitem{tilley_superfluidity_1990}
D.~R. Tilley and J.~Tilley, {\em Superfluidity and {Superconductivity}}.
\newblock CRC Press, Jan. 1990.

\bibitem{enss_low-temperature_2005}
C.~Enss and S.~Hunklinger, {\em Low-{Temperature} {Physics}}.
\newblock Springer, Apr. 2005.

\bibitem{mcauslan_microphotonic_2016}
D.~McAuslan, G.~Harris, C.~Baker, Y.~Sachkou, X.~He, E.~Sheridan, and W.~Bowen,
  ``Microphotonic {Forces} from {Superfluid} {Flow},'' {\em Physical Review X},
  vol.~6, p.~021012, Apr. 2016.

\bibitem{sachkou_coherent_2019}
Y.~P. Sachkou, C.~G. Baker, G.~I. Harris, O.~R. Stockdale, S.~Forstner, M.~T.
  Reeves, X.~He, D.~L. McAuslan, A.~S. Bradley, M.~J. Davis, and W.~P. Bowen,
  ``Coherent vortex dynamics in a strongly interacting superfluid on a silicon
  chip,'' {\em Science}, vol.~366, pp.~1480--1485, Dec. 2019.

\bibitem{schechter_observation_1998}
A.~M.~R. Schechter, R.~W. Simmonds, R.~E. Packard, and J.~C. Davis,
  ``Observation of ‘third sound’ in superfluid {3He},'' {\em Nature},
  vol.~396, pp.~554--557, Dec. 1998.

\bibitem{kashkanova_superfluid_2017}
A.~D. Kashkanova, A.~B. Shkarin, C.~D. Brown, N.~E. Flowers-Jacobs,
  L.~Childress, S.~W. Hoch, L.~Hohmann, K.~Ott, J.~Reichel, and J.~G.~E.
  Harris, ``Superfluid {Brillouin} optomechanics,'' {\em Nature Physics},
  vol.~13, pp.~74--79, Jan. 2017.

\bibitem{shkarin_quantum_2019}
A.~Shkarin, A.~Kashkanova, C.~Brown, S.~Garcia, K.~Ott, J.~Reichel, and
  J.~Harris, ``Quantum {Optomechanics} in a {Liquid},'' {\em Physical Review
  Letters}, vol.~122, p.~153601, Apr. 2019.

\bibitem{harris_proposal_2020}
G.~I. Harris, A.~Sawadsky, Y.~L. Sfendla, W.~W. Wasserman, W.~P. Bowen, and
  C.~G. Baker, ``Proposal for a quantum traveling {Brillouin} resonator,'' {\em
  Optics Express}, vol.~28, pp.~22450--22461, July 2020.
\newblock Publisher: Optical Society of America.

\bibitem{souris_ultralow-dissipation_2017}
F.~Souris, X.~Rojas, P.~Kim, and J.~Davis, ``Ultralow-{Dissipation}
  {Superfluid} {Micromechanical} {Resonator},'' {\em Physical Review Applied},
  vol.~7, p.~044008, Apr. 2017.

\bibitem{ding_high_2010}
L.~Ding, C.~Baker, P.~Senellart, A.~Lemaitre, S.~Ducci, G.~Leo, and I.~Favero,
  ``High {Frequency} {GaAs} {Nano}-{Optomechanical} {Disk} {Resonator},'' {\em
  Physical Review Letters}, vol.~105, Dec. 2010.

\bibitem{donnelly_observed_1998}
R.~J. Donnelly and C.~F. Barenghi, ``The {Observed} {Properties} of {Liquid}
  {Helium} at the {Saturated} {Vapor} {Pressure},'' {\em Journal of Physical
  and Chemical Reference Data}, vol.~27, pp.~1217--1274, Nov. 1998.

\bibitem{bowen_quantum_2015}
W.~P. Bowen and G.~J. Milburn, {\em Quantum optomechanics}.
\newblock CRC Press, 2015.

\bibitem{london_thermodynamics_1939}
H.~London, ``Thermodynamics of the thermomechanical effect of liquid {He}
  {II},'' {\em Proceedings of the Royal Society of London. Series A,
  Mathematical and Physical Sciences}, pp.~484--496, 1939.

\bibitem{sidebotham_heat_2015}
G.~Sidebotham, {\em Heat {Transfer} {Modeling}}.
\newblock Cham: Springer International Publishing, 2015.

\bibitem{swift_fundamental_2001}
G.~Swift, T.~Molinski, and W.~Lehn, ``A fundamental approach to transformer
  thermal modeling. {I}. {Theory} and equivalent circuit,'' {\em IEEE
  Transactions on Power Delivery}, vol.~16, pp.~171--175, Apr. 2001.

\bibitem{altet_thermal_2002}
J.~Altet and A.~Rubio, {\em Thermal {Testing} of {Integrated} {Circuits}}.
\newblock Boston, MA: Springer US, 2002.

\bibitem{kontoleon_dynamic_2012}
K.~J. Kontoleon, ``Dynamic thermal circuit modelling with distribution of
  internal solar radiation on varying façade orientations,'' {\em Energy and
  Buildings}, vol.~47, pp.~139--150, Apr. 2012.

\bibitem{gan_development_2020}
Y.~Gan, J.~Wang, J.~Liang, Z.~Huang, and M.~Hu, ``Development of thermal
  equivalent circuit model of heat pipe-based thermal management system for a
  battery module with cylindrical cells,'' {\em Applied Thermal Engineering},
  vol.~164, p.~114523, Jan. 2020.

\bibitem{Anderson1975}
M.~P. Zaitlin and A.~C. Anderson, ``Phonon thermal transport in noncrystalline
  materials,'' {\em Phys. Rev. B}, vol.~12, pp.~4475--4486, Nov 1975.

\bibitem{zeller_thermal_1971}
R.~C. Zeller and R.~O. Pohl, ``Thermal conductivity and specific heat of
  noncrystalline solids,'' {\em Physical Review B}, vol.~4, no.~6, p.~2029,
  1971.

\bibitem{Pollack1969}
G.~L. Pollack, ``Kapitza resistance,'' {\em Rev. Mod. Phys.}, vol.~41,
  pp.~48--81, Jan 1969.

\bibitem{long_superfluidity_1955}
E.~Long and L.~Meyer, ``Superfluidity and {Heat} {Transport} in the
  {Unsaturated} {Helium}-{II} {Film},'' {\em Physical Review}, vol.~98, no.~6,
  p.~1616, 1955.

\bibitem{atkins_third_1959}
K.~R. Atkins, ``Third and {Fourth} {Sound} in {Liquid} {Helium} {II},'' {\em
  Physical Review}, vol.~113, pp.~962--965, Feb. 1959.

\bibitem{metzger_self-induced_2008}
C.~Metzger, M.~Ludwig, C.~Neuenhahn, A.~Ortlieb, I.~Favero, K.~Karrai, and
  F.~Marquardt, ``Self-{Induced} {Oscillations} in an {Optomechanical} {System}
  {Driven} by {Bolometric} {Backaction},'' {\em Physical Review Letters},
  vol.~101, p.~133903, Sept. 2008.
\newblock Publisher: American Physical Society.

\bibitem{restrepo_classical_2011}
J.~Restrepo, J.~Gabelli, C.~Ciuti, and I.~Favero, ``Classical and quantum
  theory of photothermal cavity cooling of a mechanical oscillator,'' {\em
  Comptes Rendus Physique}, vol.~12, pp.~860--870, Dec. 2011.

\bibitem{metzger_cavity_2004}
C.~H. Metzger and K.~Karrai, ``Cavity cooling of a microlever,'' {\em Nature},
  vol.~432, no.~7020, pp.~1002--1005, 2004.

\bibitem{metzger_optical_2008}
C.~Metzger, I.~Favero, A.~Ortlieb, and K.~Karrai, ``Optical self cooling of a
  deformable {Fabry}-{Perot} cavity in the classical limit,'' {\em Physical
  Review B}, vol.~78, p.~035309, July 2008.

\bibitem{de_liberato_quantum_2011}
S.~De~Liberato, N.~Lambert, and F.~Nori, ``Quantum noise in photothermal
  cooling,'' {\em Physical Review A}, vol.~83, p.~033809, Mar. 2011.

\bibitem{aspelmeyer_cavity_2014}
M.~Aspelmeyer, T.~J. Kippenberg, and F.~Marquardt, ``Cavity optomechanics,''
  {\em Reviews of Modern Physics}, vol.~86, pp.~1391--1452, Dec. 2014.

\bibitem{haus1984waves}
H.~Haus, {\em Waves and fields in optoelectronics.}
\newblock Prentice-Hall, 1984.

\bibitem{browne1984nonlinear}
D.~Browne, ``Nonlinear effects in the damping of third-sound pulses,'' {\em
  Journal of low temperature physics}, vol.~57, no.~3, pp.~207--226, 1984.

\bibitem{meenehan_silicon_2014}
S.~M. Meenehan, J.~D. Cohen, S.~Gröblacher, J.~T. Hill, A.~H. Safavi-Naeini,
  M.~Aspelmeyer, and O.~Painter, ``Silicon optomechanical crystal resonator at
  millikelvin temperatures,'' {\em Physical Review A}, vol.~90, p.~011803, July
  2014.

\bibitem{Atkins1951}
K.~R. Atkins and C.~E. Chase, ``The velocity of first sound in liquid helium,''
  {\em Proceedings of the Physical Society. Section A}, vol.~64, pp.~826--833,
  sep 1951.

\bibitem{noauthor_comsol_nodate}
``Comsol material library.''

\bibitem{raychaudhuri_origin_1989}
A.~K. Raychaudhuri, ``Origin of the plateau in the low-temperature thermal
  conductivity of silica,'' {\em Physical Review B}, vol.~39, pp.~1927--1931,
  Jan. 1989.

\bibitem{rumble2017crc}
J.~Rumble, {\em {CRC} handbook of chemistry and physics}.
\newblock CRC Press llc Boca Raton, FL, 2017.

\bibitem{urgell_cooling_2020}
C.~Urgell, W.~Yang, S.~L. De~Bonis, C.~Samanta, M.~J. Esplandiu, Q.~Dong,
  Y.~Jin, and A.~Bachtold, ``Cooling and self-oscillation in a nanotube
  electromechanical resonator,'' {\em Nature Physics}, vol.~16, pp.~32--37,
  Jan. 2020.

\bibitem{pobell_matter_2007}
F.~Pobell, {\em Matter and methods at low temperatures}.
\newblock Berlin ; New York: Springer, 3rd, rev. and expanded ed~ed., 2007.

\bibitem{chan_laser_2011}
J.~Chan, T.~P.~M. Alegre, A.~H. Safavi-Naeini, J.~T. Hill, A.~Krause,
  S.~Gröblacher, M.~Aspelmeyer, and O.~Painter, ``Laser cooling of a
  nanomechanical oscillator into its quantum ground state,'' {\em Nature},
  vol.~478, pp.~89--92, Oct. 2011.

\bibitem{gauster_low-temperature_1971}
W.~B. Gauster, ``Low-{Temperature} {Gr{\"u}neisen} {Parameters} for {Silicon}
  and {Aluminum},'' {\em Physical Review B}, vol.~4, pp.~1288--1296, Aug. 1971.

\end{thebibliography}
\bibliographystyle{ieeetr}

\end{document}


\newcommand{\meff}{m_{\mathrm{eff}}}
\newcommand{\kappaex}{\kappa_{\mathrm{ex}}}
\newcommand{\avdw}{\alpha_{\mathrm{vdW}}}
\newcommand{\tauth}{\tau_{\mathrm{th}}}
\newcommand{\ncav}{n_{\mathrm{cav}}}
\newcommand{\alphaabs}{\alpha_{\mathrm{abs}}}
\newcommand{\epssf}{\varepsilon_{\mathrm{sf}}}
\newcommand{\rmd}{\mathrm{d}}

\title[Strong entropic interactions between light and motion]{Engineered entropic forces allow ultrastrong dynamical backaction - Supplementary Material}

\author{Andreas Sawadsky}
\author{Raymond A. Harrison}
\author{Glen I. Harris}
\author{Walter W. Wasserman}
\author{Yasmine L. Sfendla}
\author{Warwick P. Bowen}
\email{w.bowen@uq.edu.au}
\author{Christopher G. Baker}
\affiliation{ARC Centre of Excellence for Engineered Quantum Systems, School of Mathematics and Physics, University of Queensland, St Lucia, QLD 4072, Australia.}%

\date{\today}
\maketitle

\onecolumngrid

\newcommand{\beginsupplement}{%
        \setcounter{table}{0}
        \renewcommand{\thetable}{S\arabic{table}}%
        \setcounter{figure}{0}
        \renewcommand{\thefigure}{S\arabic{figure}}%
     }
\beginsupplement

\tableofcontents

\section{Experimental details}
\label{suppsectionexperimentaldetails}

\subsection{Experimental setup}

The microsphere resonator is located in a superfluid-tight sample chamber at the bottom of a
Bluefors dilution refrigerator (base temperature 10 mK)~\cite{he_strong_2020}. Telecom laser light ($\lambda=1554$ nm) from a low-noise erbium-doped fiber laser (Koheras ADJUSTIK) is evanescently coupled into the microsphere via a tapered optical fiber \cite{harris_laser_2016}. Precise fiber positioning is achieved through Attocube nanopositioning stages. The measurements are performed with the pulse-tube cooler (PTC) turned off in order to minimize vibrations. When the PTC is switched off, substitute cooling power is provided by a liquid helium `battery', containing approximately 1L of liquefied $^4$He, located on the 4K stage, providing about 2-3 hours of measurement  with the pulse tube off.  The sample chamber
contains a small volume of alumina nanoparticles in order to increase the effective chamber surface area ($\sim$10 m$^2$), leading to more precise film thickness control and greater film thickness stability~\cite{ellis_observation_1989}. While at base temperature, $^4$He gas can be injected from the top of the cryostat into the sample chamber through a thin capillary, allowing for in situ control of the superfluid film thickness.

\subsection{Superfluid film thickness}
\label{sectionsuppsuperfluidfilmthickness}

The mean superfluid film thickness $d_0$ covering the microsphere resonator can be estimated by tracking the optical resonance frequency shift $\Delta \omega_0$ that a whispering gallery mode (WGM) experiences as a superfluid film forms onto the microsphere surface~\cite{harris_laser_2016,he_strong_2020}. The film thickness $d_0$ is then given by $\Delta \omega_0/G$, with $G=\frac{\partial \omega_0}{\partial x}$  the optomechanical coupling rate which describes the optical cavity angular resonance frequency shift per unit deposited superfluid film thickness on the resonator surface \cite{baker_theoretical_2016}. Microsphere resonators support a wide variety of WGM resonances described by their radial, polar and azimuthal mode numbers along with their (TE or TM) polarization~\cite{matsko_optical_2006}. However  all these resonances have  here a similar coupling strength $G$ (see section \ref{appendixsectionG}), such that specific identification of the tracked WGM is not required.

An additional means to determine the film thickness is available for saturated films, as used in these experiments. As described above, the film is formed and thickened by injecting controlled volumes of $^4$He gas into the sample chamber via a capillary. After a certain volume, any additional injected $^4$He gas leaves the WGM resonance frequencies essentially unaffected. This corresponds to the regime of saturated film \cite{tilley_superfluidity_1990,enss_low-temperature_2005}, which differs from the unsaturated regime of our previous works~\cite{harris_laser_2016,mcauslan_microphotonic_2016,sachkou_coherent_2019}.  At this point the helium pressure in the chamber is equal to the saturated vapor pressure $p_0$, and any additional helium gas liquefies into a superfluid reservoir at the lowest point of the sample chamber. In this saturated regime, the film thickness is solely determined by the height $z$ between the microsphere and the reservoir, and can be obtained by equating the van der Waals and gravitational chemical potentials $\mu_{\mathrm{vdW}}=\frac{-\avdw}{d^3}$ and $\mu_{\mathrm{grav}}=g z$, yielding~\cite{enss_low-temperature_2005}:
\begin{equation}
    d_0=\sqrt[3]{\frac{\avdw}{g z}}
    \label{Eqfilmthicknessvsheight}
\end{equation}
Here $\avdw=2.6\times 10^{-24}$ m$^5$s$^{-2}$ is the van der Waals coefficient for silica~\cite{baker_theoretical_2016} and $g=9.8$ m.s$^{-2}$ the gravitational acceleration. The latter method is for our system the most precise technique to determine the film thickness.
\begin{itemize}
    \item The sphere is held at a height $z=2$\,cm +/- 1\,mm above the lowest point in the sample chamber. In the saturated regime, Eq. (\ref{Eqfilmthicknessvsheight}) predicts a film thickness of $d_0=23.7$\,nm +/- 0.4\,nm. 
    \item Using the WGM shift to determine the superfluid film thickness was less precise in these experiments. We observed an optical mode shift of 29\,pm after a first helium injection. Using an optomechanical coupling rate $G/2\pi=0.2$\,GHz/nm (see section \ref{appendixsectionG}) this optical shift  corresponds to a film thickness $d_0=18$\,nm. Several days later we added more helium in order to better approach the experimental set-point shown in Fig. 2 of the main text and observed a mechanical mode frequency shift of 13\,Hz, which suggest an additional film thickness of 2-3\,nm, adding up to a total film thickness of $\simeq21$\,nm. Since these measurements were done over several days and combine two different techniques, this value has a larger uncertainty.
    \item A third option to determine the film thickness is using the eigenmode simulations in COMSOL. This is achieved by fitting the measured experimental frequency of the fundamental third sound mode (72\,Hz), using the film thickness as fit parameter. This method leads to an estimated film thickness of $\sim$27\,nm. This value has an uncertainty of around 3\,nm, due to uncertainties in the exact length of the stem which defines the fundamental third sound mode (see section \ref{suppsectionthirdsoundmodescalculation}). The glue with which the stem is glued to the holder could affect the length of the stem. A deviation of 400\,$\mu$m could lead to a film thickness of 24\,nm with matching frequency of 72\,Hz. 
\end{itemize}

Based on these calculations, we determine the superfluid film thickness to be 24\,nm +/- 3\,nm, which is within the error bars and agrees with all three thickness estimation methods.  

\subsection{Calculation of the third sound modes of the microsphere resonator}
\label{suppsectionthirdsoundmodescalculation}
 Previous experimental work with superfluid third sound resonators mainly employed disk-shaped  resonators~\cite{ellis_observation_1989,schechter_observation_1998,sachkou_coherent_2019,harris_laser_2016,he_strong_2020}, for which analytical expressions (in the form of Bessel modes) exist for the third-sound resonances. While similar expressions exist for spheres (spherical harmonics), none naturally exist for the sound modes confined to the 2D outer surface of an arbitrary 3D geometry, such as the silica microsphere resonator including its supporting stem  shown in the SEM micrograph in Fig.~3(a) of the main text. To address this, we note that the superfluid helium flow in the third sound wave is considered inviscid, irrotational and incompressible\footnote{Indeed, while superfluid helium is in fact quite compressible \cite{kashkanova_superfluid_2017,shkarin_quantum_2019, harris_proposal_2020} (with a bulk modulus of approximately 8 MPa compared to 2 GPa for water), the van der Waals pressure exerted on the superfluid helium's film free surface (typically in the kPa range for the film thicknesses considered here~\cite{he_strong_2020}) is approximately three orders of magnitude lower than helium's bulk modulus. As a consequence, any local influx of superfluid predominantly leads to a thickening of the film and not an increase in density, such that the superfluid may be well approximated as incompressible in the third sound wave.}. As such it is a potential flow and, in the limit of small wave amplitude, the out-of-plane deflection of the superfluid surface $\eta\left(\vec{r},t\right)$ obeys the simple wave equation:
\begin{equation}
    \left(\nabla^2-\frac{1}{c^2}\frac{\partial^2}{\partial t^2}\right) \eta=0
\end{equation}
Here $c$ is the speed of sound, which neglecting the influence of surface tension, takes the form
$c_3=\sqrt{3 \frac{\rho_s}{\rho}\frac{\avdw}{d^3}}$ \cite{baker_theoretical_2016}. Assuming a separable time-harmonic standing wave solution, of the kind $\eta\left(\vec{r},t\right)=\eta\left(\vec{r}\right) e^{i\Omega t}$, leads to the Helmholtz equation for the spatial mode profile $\eta(\vec{r})$:
\begin{equation}
    \left(\nabla^2+k^2\right)\eta(\vec{r})=0, 
\end{equation}
where $k^2=\frac{\Omega^2}{c^2}$ and the displacement profile $\eta(\vec{r})$ is defined on the (2D) surface of the (3D) resonator geometry. When the resonator is a sphere of radius R, the third sound modes are given by the eigenfunctions of the angular part of the Laplacian operator $\Delta=\nabla^2$, called the spherical harmonic functions $Y^l_m\left(\theta,\phi\right)$ of degree $l$ and order $m$,  with eigenvalue $k^2=l(l+1)/R^2$ and frequency:
\begin{equation}
    \Omega=k c=\frac{c \sqrt{l(l+1)}}{R}. 
    \label{Eqfrequencysphericalharmonics}
\end{equation}
\begin{figure}
    \centering
    \includegraphics[width=\textwidth]{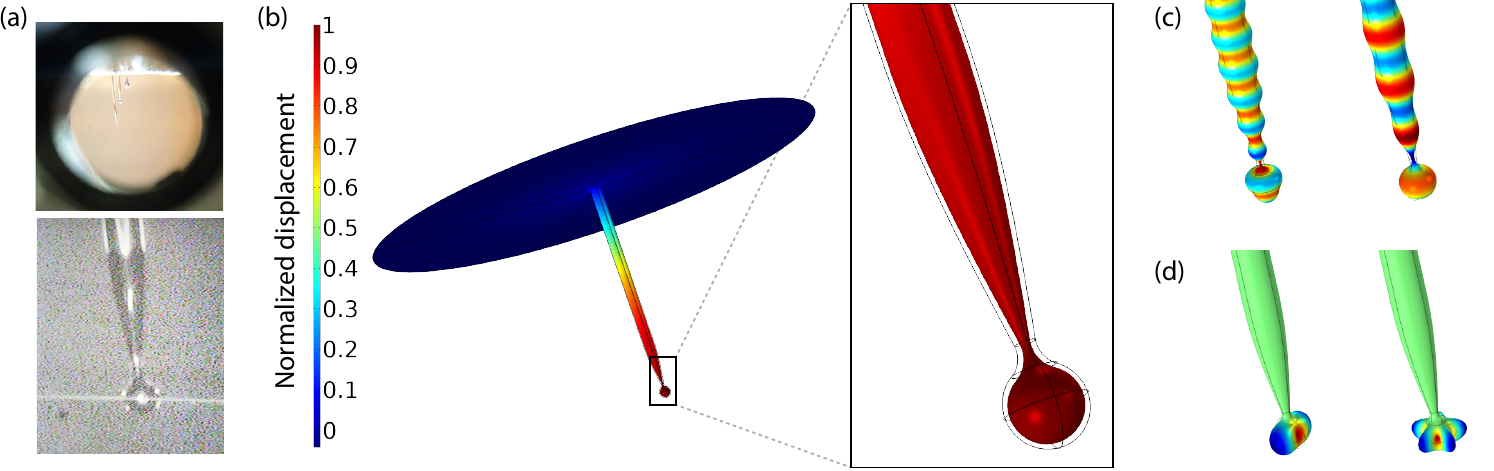}
    \caption{(a) Top: three microspheres of varying diameter and stem length imaged through the sample chamber window. Bottom: image of the silica microsphere and the coupling tapered fiber, measured at cryogenic temperature with a long working distance microscope objective. (b) Fundamental third sound eigenmode localized to the fiber stem, obtained through finite element simulations. The rapid enlargement of the cross-section represents the point at which the fiber stem is glued to the sample holder. Inset shows how the fluid motion alternatively thins (and thickens) the film around the tip of the sphere.  (c) Higher order third sound excitations of the fiber stem and microsphere. (d) Third sound modes localized to the microsphere tip, closely resembling the  $Y^2_2\left(\theta,\phi\right)$ and $Y^5_5\left(\theta,\phi\right)$ eigenmodes of an ideal sphere. }
    \label{figure_supp_modes}
\end{figure}
Similarly, the third sound modes confined to the surface of an arbitrary three dimensional geometry may be obtained through solving the Helmholtz equation on the  exterior 2D surface of this 3D geometry with the help of finite element modelling software (Comsol Multiphysics). We use this technique to obtain the modes of oscillation of a superfluid film confined to the surface of a silica microsphere whispering gallery mode resonator, including its supporting stem. The microsphere resonator is obtained by melting the end of a silica single mode fiber (SMF-28) in a fusion splicer. The non-reflown end of the fiber is then held in place on a sample holder inside the cryostat  by a large drop of UV glue (not visible here). Three such resonators, of differing stem length $l$, are shown in the top panel of Fig. \ref{figure_supp_modes}(a), protruding from the sample holder in order to allow optical access through a tapered fiber (see bottom panel). 
Fig. \ref{figure_supp_modes}(b) shows the fundamental mode of oscillation of a superfluid film confined to the surface of such a silica microsphere resonator.  Acoustic confinement is provided by the large change in acoustic impedance at the contact point to the sample holder due to the rapid change in cross-sectional area, much like in a Helmholtz resonator~\cite{souris_ultralow-dissipation_2017}. This is evidenced by the fact that
the obtained resonance frequency $\Omega_M/2\pi=\sim86$\,Hz for a 24 nanometer thick film---which reasonably closely matches that observed in the experiments---is essentially independent of the choice of  fixed (Dirichlet) or free (von Neumann) boundary condition at the edge of the simulation domain~\cite{baker_theoretical_2016}. For this fundamental acoustic resonance, superfluid oscillates back and forth between  the surrounding bath and the tip of the sphere, thereby efficiently modulating the whispering gallery mode optical path-length, as shown in the inset of Fig. \ref{figure_supp_modes}(b). Higher order excitations of the stem and sphere are displayed in Fig. \ref{figure_supp_modes} (c).

This acoustic confinement through impedance mismatch is also at play at the level of the thin neck which joins the silica microsphere to the silica fiber stem. This results in (higher frequency) third sound modes localized on the spherical tip, as shown in Fig. \ref{figure_supp_modes}(d), with mode profiles and eigenfrequencies closely matching those given by the spherical harmonic functions $Y^l_m\left(\theta,\phi\right)$ describing the eigenmodes of a perfect sphere (see section below).

\begin{figure}
    \centering
    \includegraphics[width=0.6\textwidth]{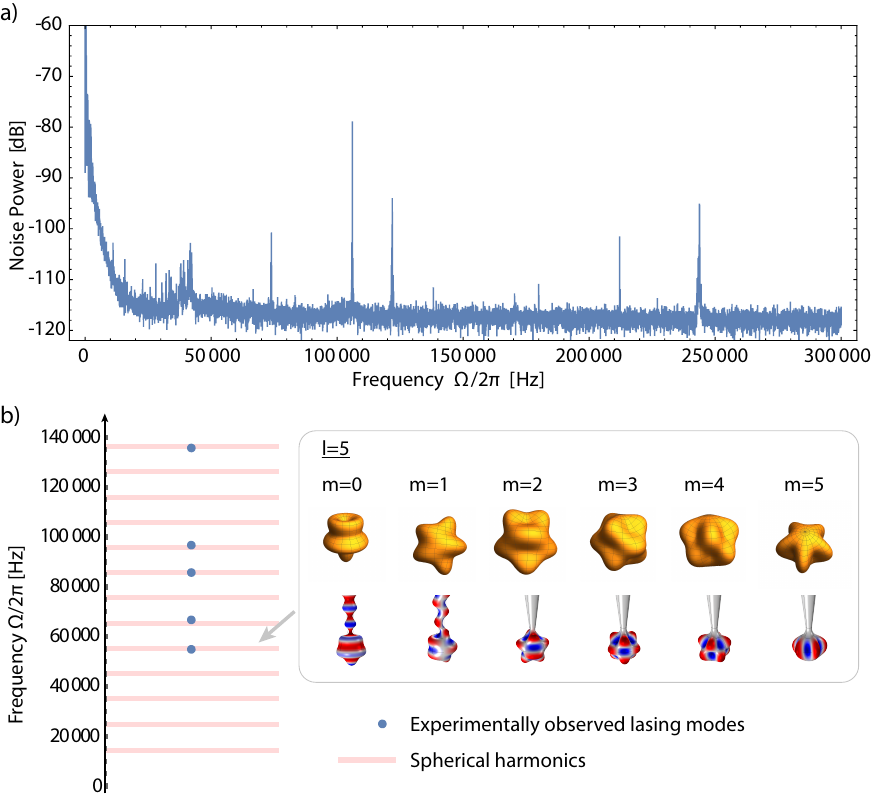}
    \caption{(a) Power spectrum showing a number of high-frequency third sound modes, acquired with a film thickness of approximately 7.5\,nm. (b) Matching of experimentally observed lasing modes (blue dots) to frequency of spherical harmonics $Y^l_m$ of an ideal sphere (see Eq. (\ref{Eqfrequencysphericalharmonics})),  (width of band 2\,kHz). }
    \label{figspheremodessup}
\end{figure}

In addition to the fundamental stem mode discussed in the main text, we observe a number of high-frequency modes consistent with third sound modes localized on the microsphere itself. Fig. \ref{figspheremodessup} shows a representative spectrum, acquired with a film thickness of $\sim 7.5$ nm. A number of third sound modes are visible with frequencies ranging from tens to hundreds of kHz and  Q factors in the $10^4$ range. These modes can be brought into regenerative oscillation with nanowatts of optical power. Which particular mode experiences dynamical backaction is strongly dependent on laser-cavity detuning. Fig. \ref{figspheremodessup}(b)  records the frequencies of the third sound modes which could be brought into lasing during an experimental run. Both their  density and frequency are consistent with spherical harmonics $Y^l_m$ of an ideal sphere (pink bands, where $l$ is incremented from 1 to 13). More precise mode identification was not performed here, as identifying the mode frequency provides information only on the degree $l$ (see Eq. \ref{Eqfrequencysphericalharmonics}), and  spherical harmonics of degree $l$ have $2l+1$ degeneracy (the order  m can take integer values from $-l$ to $l$). This is illustrated in the inset of Fig. \ref{figspheremodessup}(b), which displays the analytical spherical harmonic modes of an ideal sphere $Y^l_m\left(\theta,\phi\right)$, along with the corresponding eigenmodes of the sphere with stem obtained through finite element simulation (negative values of m which rotate the eigenmode are not shown here).
\subsubsection*{Sphere modes}
\label{sectionspheremodesYlm}

\subsection{ Calculation of effective mass, coupling rate and thermal conductance}

\subsubsection*{Optomechanical coupling G}
\label{appendixsectionG}
\begin{figure}
    \centering
    \includegraphics[width=.9\textwidth]{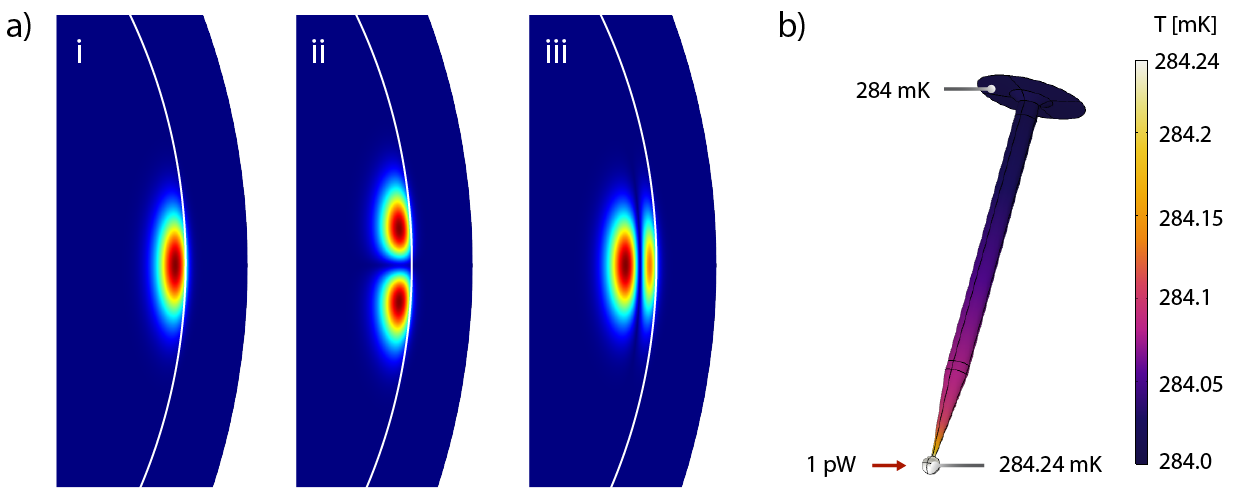}
    \caption{(a) Whispering gallery modes of a 55\, $\mu$m radius microsphere resonator, obtained through finite element simulation. i) Fundamental radial $n=1$ and polar $l=1$ WGM  with azimuthal number $m=320$ and resonance wavelength $\lambda_0=1506$ nm. Its optomechanical coupling strength calculated through FEM~\cite{baker_theoretical_2016} is $G/2\pi=1.883\times 10^{17}$ Hz/m. ii) Higher order polar WGM ($n=1$; $l=2$; $m=320$), with optomechanical coupling strength $G/2\pi=1.889\times 10^{17}$ Hz/m. iii) Higher order radial WGM ($n=2$; $l=1$; $m=320$), with  $G/2\pi=1.903\times 10^{17}$ Hz/m.  (b) Calculation of the thermal conductance of the microsphere resonator. An  absorbed power of 1\,pW at the level of the spherical tip leads to a steady-state temperature increase by 0.24\,mK, corresponding to a thermal conductance  $G_{\mathrm{th}}=4.1\times 10^{-9}$ W/K at a thermal bath temperature of 284\,mK. Physical parameters used in the simulation are summarized in Table \ref{Table_physical_params}. }
    \label{FigWGMs_temp}
\end{figure}

Calculating the optomechanical coupling $G$ requires identification of the employed WGM in order to compute its field overlap with the superfluid coating the resonator~\cite{baker_theoretical_2016}.  However, microspheres have a very dense whispering gallery mode spectrum, with WGMs differing by their radial, polar and azimuthal mode orders ($n$, $l$ and $m$ respectively), along with their TE or TM polarization \cite{matsko_optical_2006}, as illustrated in Fig. \ref{FigWGMs_temp}(a). This large WGM mode density makes it difficult to identify the mode used in the experiments. Fortunately, changes in the WGM order have only a modest influence on the coupling strength, with a $<1$\% change arising from incrementing the radial or polar order beyond the fundamental mode (see Fig. \ref{FigWGMs_temp}(a)).

A larger difference, on the order of 5\%, arises between TE and TM polarizations. Indeed, the WGMs with a dominant radial E field component have a larger field at the surface due to the orthogonal E field discontinuity at the silica interface~\cite{baker_theoretical_2016}. Calculation for 100 WGMs closest in resonance wavelength to 1550 nm shows that their G is bounded between $1.9\times 10^{17}$ and $2.1\times 10^{17}$ Hz/m, allowing us to constrain the uncertainty to within $\sim$10\%. These calculated values are in good agreement with the analytical expression for a circular WGM resonator $G=-\frac{\omega_0}{R}$~\cite{ding_high_2010}, corrected for the lower dielectric permittivity of superfluid helium~\cite{baker_theoretical_2016}:
\begin{equation}
    G=\frac{\partial \omega_0}{\partial x}\simeq-\frac{\omega_0}{R}\left(\frac{1-\varepsilon_{\mathrm{sf}}}{1-\varepsilon_{\mathrm{SiO_2}}}\right), 
\end{equation}
which predicts $G/2\pi=1.93\times 10^{17}$\,Hz/m  for a 55\,$\mu$m radius sphere. Here $\varepsilon_{\mathrm{sf}}=1.058$ is the relative permittivity of superfluid helium~\cite{donnelly_observed_1998}, and $\varepsilon_{\mathrm{SiO_2}}=2.1$ that of silica. We note the value of $G$ is calculated with a radius of 55 $\mu$m, obtained by an optical microscope measurement. Elsewhere a value of 49.5 $\mu$m obtained by SEM is used. This discrepancy is due to the oblate shape of the reflown microsphere.

\subsubsection*{Effective mass}
\label{appendixsectionmeff}

The effective mass of an acoustic mode taken at a reduction point $\vec{R}$
is obtained by reducing the system to a point mass $\meff$ moving with velocity $v(\vec{R})$ possessing the same kinetic energy $E_k$ as the original system, that is $\meff=\frac{2 E_k}{v(\vec{R})^2}$. For a third sound mode,  this takes the form~\cite{baker_theoretical_2016}:
\begin{equation}
    m_{\mathrm{eff}}=\frac{2 E_p}{v^2(\vec{R})}=\frac{2\iint \frac{3 \rho \alpha_{\mathrm{vdW}} \eta^2\left(\vec{r}\right) \mathrm{d}^2(\vec{r}) }{2 d^4}}{\eta^2(\vec{R})\,\Omega^2},
\end{equation}
where $E_p=E_k$ is the potential energy stored in the third sound wave, and the integral is taken over the surface $\mathcal{A}$ of the resonator.  For the fundamental mode of the sphere and stem
shown in Fig. \ref{figure_supp_modes}(b), with a reduction point on the equator of the microsphere and a 24\,nm film thickness, 
$\meff=5.1 \times 10^{-3}$\,kg. Note that this value is approximately 2 billion times larger than the total mass of superfluid covering the resonator $m=\mathcal{A}\, d\, \rho_{\mathrm{He}}=2.65\times 10^{-12}$\,kg. The larger effective mass arises from the fact we consider here only the out-of-plane displacement $\eta$ of the fluid interface (which couples to the light), while the majority of the superflow occurs in plane~\cite{baker_theoretical_2016}.

\subsubsection*{\texorpdfstring{Radiation pressure Single photon optomechanical coupling rate $g_{0_{rp}}$}{Radiation pressure Single photon optomechanical coupling rate g\_0\_rp}}
For the low-frequency stem mode shown in Fig. \ref{figure_supp_modes}(b), the superfluid displacement is uniform along the tip of the microsphere where the light is confined, such that the radiation pressure single photon optomechanical coupling rate $g_{0_{rp}}$ is given by~\cite{baker_theoretical_2016, bowen_quantum_2015}:
\begin{equation}
    g_{0_{rp}}=G x_{\mathrm{zpf}} =G \sqrt{\frac{\hbar}{2 \meff \Omega}}.
\end{equation}
With $G/2\pi= 2 \times  10^{17}$ Hz/m (see section \ref{appendixsectionG}) and $\meff=5.1 \times 10^{-3}$ kg (see section \ref{appendixsectionmeff}), this yields $x_{\mathrm{zpf}}= 4.8\times 10^{-18}$ m and $g_{0_{rp}}/2\pi=0.95$ Hz.

\section{Thermal-electric circuit analogy}
\label{suppsectionthermalequivalentcircuit}
The fountain pressure in superfluid helium is given by~\cite{london_thermodynamics_1939}:
\begin{equation}
    P_{\mathrm{fp}}=\rho_{\mathrm{He}}\, S_{\mathrm{He}}(T)\, \Delta T_{\mathrm{He}},
    \label{Eq:fountain_pressure_SF}
\end{equation}

where $\rho_{\mathrm{He}}$ is the superfluid helium density, $S{_\mathrm{He}}(T)$ is the temperature-dependent entropy of helium, and  $\Delta T_{\mathrm{He}}=T_{\mathrm{He}}-T$ is the  difference between the environment  temperature $T$ and the superfluid film covering the resonator at temperature $T_{\mathrm{He}}$. When calculating the fountain pressure force the challenge is to precisely estimate the temperature rise in the superfluid film, because it strongly depends on the thermal parameters of the system (thermal conductivity, specific heat, Kapitza resistance, vapor pressure etc), which in turn are all strongly temperature-dependent. To model this system we use the  technique of the thermal-electric analogy.  

The thermal-electric analogy as a lumped-element model is a well known approach to analyse and simulate a variety of complex thermal systems~\cite{sidebotham_heat_2015, swift_fundamental_2001}. Applications of thermal equivalent circuits range from designing heat sinks for semiconductor circuits~\cite{altet_thermal_2002} and understanding the impact of solar radiation on building energy consumption~\cite{kontoleon_dynamic_2012} to battery pack thermal management~\cite{gan_development_2020}. Here we use the analogy between thermal quantities and electric quantities (summarized in Table \ref{tab:Table_thermal_electric_analogy}) to transform our thermal system into an electric circuit analog. 
\begin{figure}
    \centering
    \includegraphics[width=0.9\textwidth]{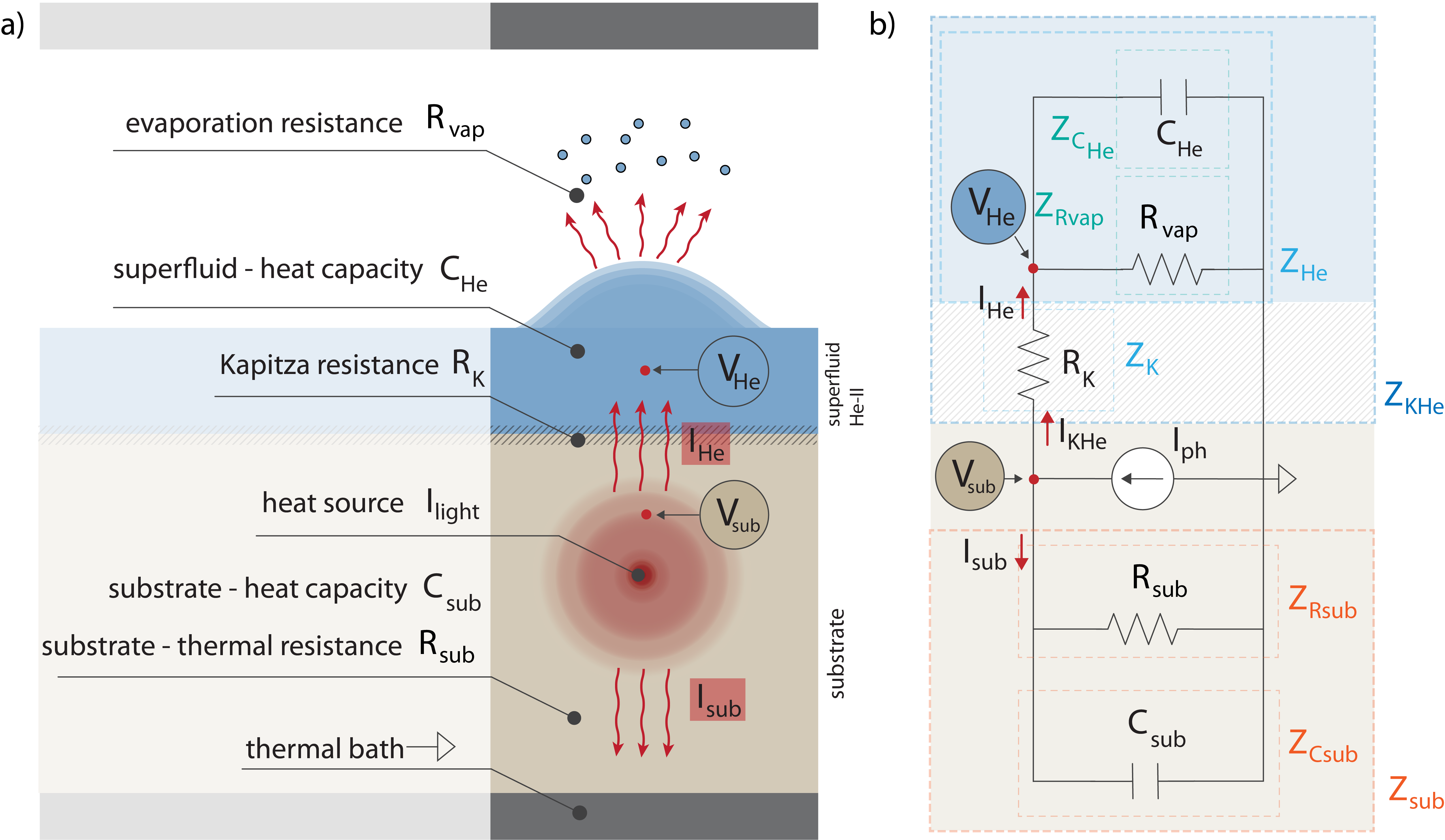}
    \caption{\textbf{Electric-thermal analogy scheme.} a) shows the schematic of our thermal system with the analog electric quantities. The analogy for the heat source (absorbed intra cavity power) is a current source $I_{\mathrm{light}}$ and the electric ground represents the thermal bath of the system.  b) Electric circuit analogy of our thermal system. The voltages $V_{\mathrm{sub}}$ and $V_{\mathrm{He}}$ are the analogs of the temperatures in the substrate and the superfluid film.}
    \label{fig:electric_circuit}
\end{figure}
In figure \ref{fig:electric_circuit} a) we show a schematic of our thermal system with the equivalent electric quantities and in b) the electric circuit as an analog representation of our thermal system. The heat source in our system is the absorbed intracavity optical power in the silica sphere, which is represented by the current source $I_{\mathrm{ph}}$ in the circuit. There are two paths for the heat flow towards the thermal bath (dark grey). First it can flow through the substrate itself, i.e. the silica stem which is thermally anchored to the cryostat (beige colored area). Second it may flow through the silica/superfluid interface (striped line) with interfacial resistance $R_K$ into the superfluid helium (light blue color) and dissipate via evaporation. Because the heat can dissipate through both paths simultaneously, these are arranged in parallel in the electric circuit. The two main quantities we are interested in are the temperature changes in the substrate $\Delta T_{\mathrm{sub}}$ and in the superfluid helium thin film $\Delta T_{\mathrm{He}}$ with respect to the thermal bath. These two quantities are represented by the two thermal potentials (electric analogy: voltage) $V_{\mathrm{sub}}$ and $V_{\mathrm{He}}$ with respect to some reference voltage, which is the thermal bath (electric analogy: ground). They depend on the thermal resistances $R_{\mathrm{th}}$, capacitance $C_{\mathrm{th}}$, thermal bath temperature $T$ and heat flow rate given by the absorb photons, which is the current $I_{\mathrm{ph}}$ in the electric analogy. Knowing these values, we eventually can calculate the fountain pressure and force. 

\begin{table}
    \centering
    \begin{tabular}{c| c |c}
    \hline
      electric & thermal &  superfluid mass flow \\
      \hline\hline
        charge $q [A \cdot s]$ & heat $Q [J]$ &  mass $m [kg]$\\
        current $I [A]$ & heat flow rate  $\dot Q [W]$ &  mass flow rate $\dot m [kg/s]$\\
        voltage $V [V]$ & temperature $T [K]$ &  chemical potential $\mu [J/kg]$\\
        resistance $R [V/A]$ & heat resistance  $ R_{\mathrm{th}} [K/W]$ &  -\\
        capacitance $C[A \cdot s/V] $ & heat capacitance $C_{\mathrm{th}} [J/K]$ &   mass capacitance $C_{\mathrm{m}}=dm/d\mu [1/J]$\\
        inductance $L [H]$ & - & mass inductance $L_{\mathrm{m}} [m^{2}/kg] $\\
        \hline
         $I=\Delta V/R $ & $\dot Q = \Delta T/R_{\mathrm{th}}$ &  -\\
        $I=C dV/dt $ & $\dot Q = C_{\mathrm{th}} dT/dt$ &   $\dot m = C_{\mathrm{m}} d\mu/dt$\\
        Kirchoff's current law & first law of thermodynamics &  law of conservation of mass\\
    \end{tabular}
    \caption{Electric, thermal and superfluid mass flow analog quantities.}
    \label{tab:Table_thermal_electric_analogy}
\end{table}

\subsection{Thermal-electric elements of substrate - silica}
In a lumped-element model for heat transfer an  element with a non-zero heat capacity is is modeled by two quantities in a parallel configuration: a thermal resistance $R_{\mathrm{th-sub}}$ and a thermal capacitance  $C_{\mathrm{th-sub}}$ which accounts for the element's thermal inertia. 

The thermal resistance $R_{\mathrm{sub}}$ can be simulated with COMSOL. All parameters used in this simulation are provided in Table \ref{Table_physical_params} and are for the bath temperature of 284\,mK.
Fig. \ref{FigWGMs_temp}(b) shows the steady-state temperature increase in the microsphere plus stem system with $\dot{Q}=1$ pW of optical power absorbed at the level of the microsphere. The 0.24\,mK temperature increase corresponds to a thermal resistance $R_{\mathrm{th-sub}}=2.44\times10^8$\,K/W and a thermal conductance $G_{\mathrm{th-sub}}= R_{\mathrm{th-sub}}^{-1}=\kappa_{\mathrm{SiO_2 @284mK}}\times\phi = 4.1\times 10^{-9}$\,W/K, where $\kappa_{\mathrm{SiO_2 @284\,mK}}=1.6\times10^{-3}$\,W/m/K is the thermal conductivity for silica at 284\,mK and $\phi =2.52\times 10^{-6}$\,m is a temperature-independent geometrical factor (which we obtain from this simulation). The thermal conductivity of silica $\kappa_{\mathrm{SiO_2}}(T)$ is temperature dependent. In figure \ref{fig:temp-parameters} a) we fitted a function of the form $\kappa_{\mathrm{SiO_2}}(T)=0.0000488638\times T + 0.0212904\times T^2 - 0.00436582\times T^3 - 0.000101651\times T^4$ to the data from \cite{Anderson1975} to get a temperature-dependent thermal resistance of the silica microsphere  $R_{\mathrm{th-sub}}(T)=(\kappa_{\mathrm{SiO_2}}(T)\times \phi)^{-1}$.

The thermal capacitance of the substrate is given by $C_{\mathrm{th-sub}}= c_{\mathrm{SiO_2}}(T)\times m_\mathrm{sub}$, with $m_\mathrm{sub}$ being the mass of the microsphere (incl. stem) and $c_{\mathrm{SiO_2}}(T)= 0.00105\times T + 0.0018\times T^3$ is the temperature-dependent specific heat capacity for silica (see figure \ref{fig:temp-parameters} b)), which is a fit function to the data from \cite{zeller_thermal_1971}.

\subsection{Thermal Kapitza resistance at the interface}

The interface between the silica and the superfluid thin-film results in an interfacial thermal resistance called the Kapitza resistance. It is temperature and material dependent, and arises due to the large acoustic impedance mismatch between silica and superfluid helium, reducing phonon propagation from one medium to the other. According to \cite{Pollack1969} the Kapitza resistance has the functional form:

\begin{equation}
   R^{'}_{\mathrm{K}}(T) = \frac{15 \hbar^3\rho_{\mathrm{SiO_2}} c^3_{t_{\mathrm{SiO_2}}}}{2\pi^2 k^4_{\mathrm{B}}\rho_{\mathrm{He}}c_{1_{\mathrm{He}}}F(\frac{c_{l_{\mathrm{SiO_2}}}}{c_{t_{\mathrm{SiO_2}}}})T^3}, 
\end{equation}
where $\rho_{\mathrm{SiO_2}}$ and $\rho_{\mathrm{He}}$ are the densities of silica and liquid helium, $c_{t_{\mathrm{SiO_2}}}$ and $c_{l_{\mathrm{SiO_2}}}$ are respectively  the transverse and longitudinal sound velocities of silica, $c_{1_{\mathrm{He}}}$ is the first sound speed in superfluid helium and  $F(c_{l_{\mathrm{SiO_2}}}/c_{t_{\mathrm{SiO_2}}}) = 2.5$ \cite{Pollack1969} is a silica specified function. Fig. \ref{fig:temp-parameters}(c) shows the temperature dependency of the Kapitza resistance for a silica and  superfluid helium interface with units [$\mathrm{m^2 K/W}$]. When calculating the Kapitza interfacial thermal resistance in our system $R_{\mathrm{K}}$, we need to normalize it to the silica microsphere surface area, i.e. $R_{\mathrm{K}}(T) =R^{'}_{\mathrm{K}}(T)/\mathcal{A}$ [K/W].

\subsection{Thermal-electric elements of superfluid helium He-II}

\begin{figure}
    \centering
    \includegraphics[width=0.8\textwidth]{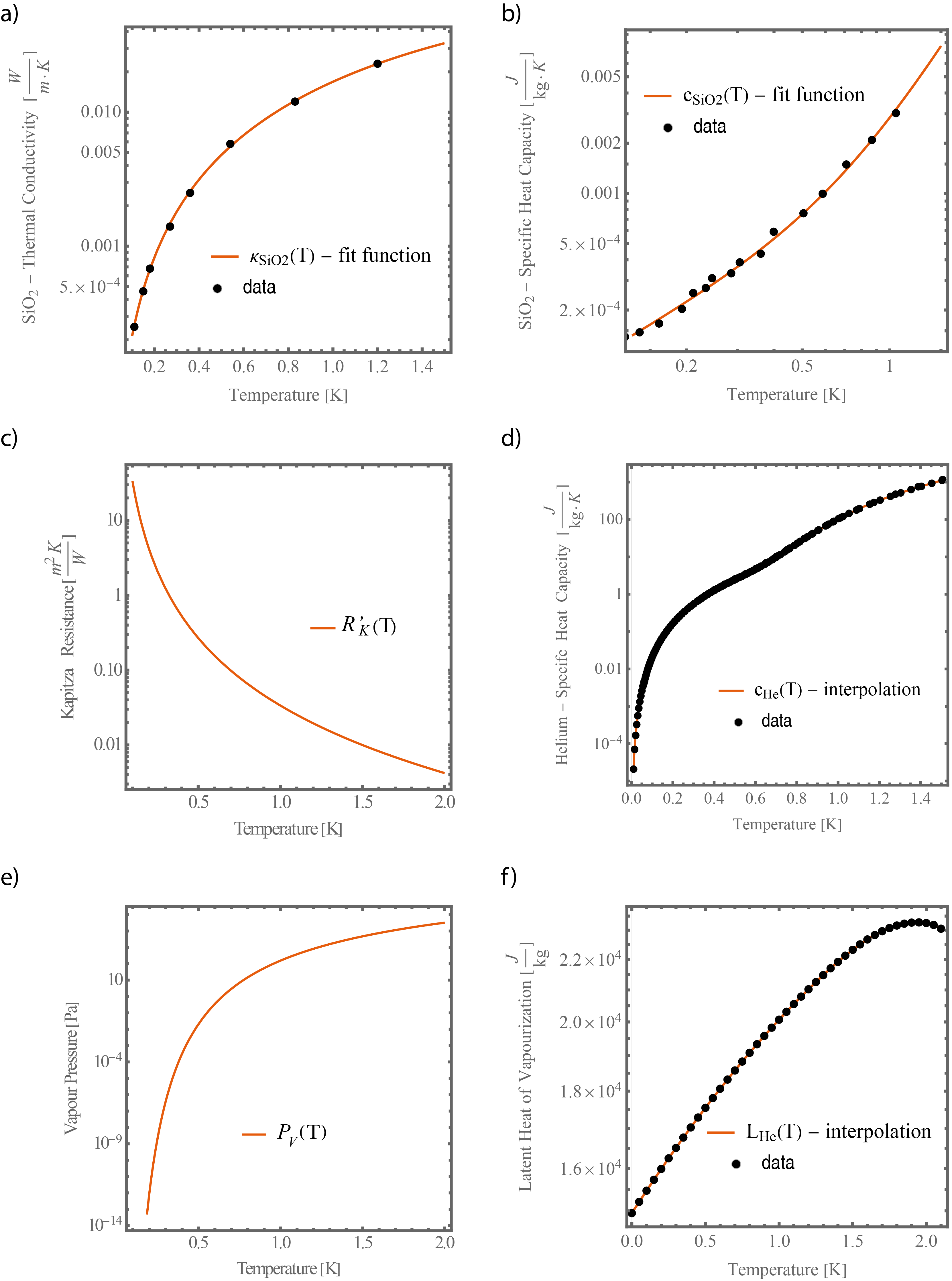}
    \caption{\textbf{Temperature-dependent material parameters.} Here we present all the temperature-dependent material parameters used for the thermal-electric analogy model. a) Fitted thermal conductivity $\kappa_{\mathrm{SiO}_{\mathrm{2}}}(T)$ of silica with data points taken from literature \cite{Anderson1975}. b) Fitted specific heat capacity $c_{\mathrm{SiO}_{\mathrm{2}}}(T)$ with measured data from \cite{zeller_thermal_1971}. c) Kapitza resistance of the interface between liquid He-II and silica \cite{Pollack1969}. The figures d),e) and f) are all helium-related parameters. In  d),  we show the interpolated the specific heat capacity of helium $c_{\mathrm{He}}(T)$ obtained from the data in Ref.~\cite{donnelly_observed_1998}.  e) plots the vapour pressure for helium, which is given in Ref.~\cite{donnelly_observed_1998} in functional form (Eq.~\ref{Eq:vapour-pressure}). In f) $L_{\mathrm{He}}(T)$ is the interpolation of the data for the latent heat of vaporization \cite{donnelly_observed_1998}.}
    \label{fig:temp-parameters}
\end{figure}

The  superfluid helium thin-film is represented by the thermal resistance  $R_{\mathrm{vap}}$  and the thermal capacitance $C_{\mathrm{He}}$. 
In thin superfluid films, the normal fluid component is viscously clamped to the substrate and does not flow. Only the superfluid component, which carries no entropy, is free to move. Thermal conductance through the liquid itself is therefore negligible, and the thermal conductivity occurs primarily through influx of superfluid, which evaporates extracting the latent heat of vaporization~\cite{long_superfluidity_1955}. To calculate the thermal conductance  $G_{\mathrm{vap}}(T)$ resulting from this evaporative process at a temperature T, we need to multiply the resulting net helium mass flow rate per unit area $\dot{m}_{\mathrm{He}}$ by the latent heat of vaporization of helium $L_{\mathrm{He}(T)}$ and divide by the  temperature change $(T_0-T)$, which gives the area-normalized `net' energy leaving due to evaporation. Multiplying by the silica microsphere area $\mathcal{A}$ gives the total evaporative thermal conductance of our superfluid film:
\begin{equation}
   G_{\mathrm{vap}}(T) = \dot{m}_{\mathrm{He}}\frac{L_{\mathrm{He}}(T)}{(T_0-T)}\mathcal{A}.
   \label{Eq:vapour-conductance}
\end{equation}
The net mass flow rate per unit area is given by \cite{atkins_third_1959}:
\begin{equation}
   \dot{m}_{\mathrm{He}}(T) = \gamma \sqrt{\frac{m_{\mathrm{He_{mol}}}}{2\pi R\, T}}\left(\frac{dP_{\mathrm{V}}}{dT}\right)_{\mathrm{v.p.c}}(T_0-T),
   \label{Eq:mass-flow}
\end{equation}
with $\gamma=1$, $m_{\mathrm{He_{mol}}}$ as the molar mass of helium, $R=8.3145$ J mol$^{-1}$ K$^{-1}$   the ideal gas constant and $(\frac{dP_{\mathrm{V}}}{dT})_{\mathrm{v.p.c}}$ the gradient of the vapour pressure curve for helium.  The saturated vapour pressure $ P_{\mathrm{V}}(T)$ curve  is plotted in Fig. \ref{fig:temp-parameters}(e), and given by \cite{donnelly_observed_1998}:
\begin{equation}
   P_{\mathrm{V}}(T) =\exp(i_o-\frac{L_0}{RT}+\frac{5}{2}\log(T)),
   \label{Eq:vapour-pressure}
\end{equation}
where $i_0=12.2440$ and $L_0=59.83$ J/mol is the latent heat of vaporization at absolute zero. The temperature-dependent latent heat of vaporization of helium $L_{\mathrm{He}}(T)$ is an interpolation function, shown in Fig.~\ref{fig:temp-parameters} f), to the data extracted from \cite{donnelly_observed_1998}. The equations \ref{Eq:vapour-conductance},\ref{Eq:mass-flow},\ref{Eq:vapour-pressure} all together enable us to derive a temperature-dependent expression for the thermal resistance via evaporation $R_{\mathrm{vap}}(T)=(G_{\mathrm{vap}}(T))^{-1}$.

The temperature-dependent heat capacity of thin-film superfluid helium is given by $C_{\mathrm{He}}(T)= c_{\mathrm{He}}(T)m_{\mathrm{He}}$, with $m_{\mathrm{He}}$ being the superfluid helium mass covering the full microsphere (incl. stem) and $c_{\mathrm{He}}(T)$ an interpolation of the temperature-dependent specific heat for superfluid helium, with data obtained from \cite{donnelly_observed_1998} and shown in figure \ref{fig:temp-parameters} d).

\subsection{\texorpdfstring{Transfer-function for the superfluid helium temperature $V_{\mathrm{He}}(\Omega,T)$}{Transfer-function for the superfluid helium temperature V\_He(Omega,T)}}
\label{voltageassuperfluidheliumtemperature}
Having specified the electric analog of each thermal quantity in our system (see sections above) and simplified it to an electric circuit (see figure \ref{fig:electric_circuit}), enables us now to use simple electric calculation techniques to determine the bath-temperature dependent superfluid helium temperature $V_{\mathrm{He}}(\Omega,T)$ and its frequency response (transfer-function) to a fluctuating heat source. This gives us a full framework of our system so we can operate at the ideal temperature and frequency to maximise and control the fountain pressure backaction.

\begin{table}
    \centering
    \begin{tabular}{p{5 cm} l l l }
    \hline
      thermal-electric quantity   & functional form & frequency domain\\
      \hline
      substrate thermal resistance & $R_{\mathrm{sub}}(T)=(\kappa_{\mathrm{SiO_2}}(T) \phi)^{-1}$ & $Z_{\mathrm{R_{sub}}}(T)=R_{\mathrm{sub}}(T)$ \\
      substrate thermal capacitance & $C_{\mathrm{sub}}(T)=c_{\mathrm{SiO_2}}(T)m_{\mathrm{sub}}$ & $\tilde{Z}_{\mathrm{C_{sub}}}(\Omega,T)=\frac{1}{i \Omega C_{\mathrm{sub}}(T)}$ \\
      Kapiza Resistance & $R_{\mathrm{K}}(T)\simeq\frac{3 \hbar^3\rho_{\mathrm{SiO_2}} c^3_{t_{\mathrm{SiO_2}}}}{\pi^2 k^4_{\mathrm{B}}\rho_{\mathrm{He}}c_{1_{\mathrm{He}}}T^3\mathcal{A}}$ & $Z_{\mathrm{K}}(T)=R_{\mathrm{K}}(T)$ \\
      superfluid He thermal resistance & $R_{\mathrm{vap}}(T)=\left(\gamma \sqrt{\frac{m_{\mathrm{He_{mol}}}}{2\pi R T}}(\frac{dP_{\mathrm{V}}}{dT})_{\mathrm{v.p.c}} L_{\mathrm{He}}(T)\mathcal{A}\right)^{-1}$ & $Z_{\mathrm{R_{vap}}}(T) = R_{\mathrm{vap}}(T)$ \\
      superfluid He thermal capacitance & $C_{\mathrm{He}}(T)= c_{\mathrm{He}}(T)m_{\mathrm{He}}$ & $\tilde{Z}_{\mathrm{C_{He}}}(\Omega,T)=\frac{1}{i\Omega C_{\mathrm{He}}(T)}$ \\
         \hline
    \end{tabular}
    \caption{Thermal-electric quantities in DC and frequency domain.}
    \label{table:impedance-resistance}
\end{table}

First we transform all thermal-electric quantities into the frequency domain, so they can be written as complex impedances, which are all summarized in table \ref{table:impedance-resistance}. The total complex impedance for the substrate and superfluid helium are respectively given by:
\begin{equation}
   \tilde{Z}_{\mathrm{sub}}(\Omega,T) =\left( \frac{1}{Z_{\mathrm{R_{sub}}}(T)}+\frac{1}{\tilde{Z}_{\mathrm{C_{sub}}}(\Omega,T)}\right)^{-1}
   \label{Eq:imp_sub}
\end{equation}
and
\begin{equation}
   \tilde{Z}_{\mathrm{He}}(\Omega,T) =\left( \frac{1}{Z_{\mathrm{R_{vap}}}(T)}+\frac{1}{\tilde{Z}_{\mathrm{C_{He}}}(\Omega,T)}\right)^{-1}.
   \label{Eq:imp_sf}
\end{equation}
\begin{figure}
    \centering
    \includegraphics[width=0.8\textwidth]{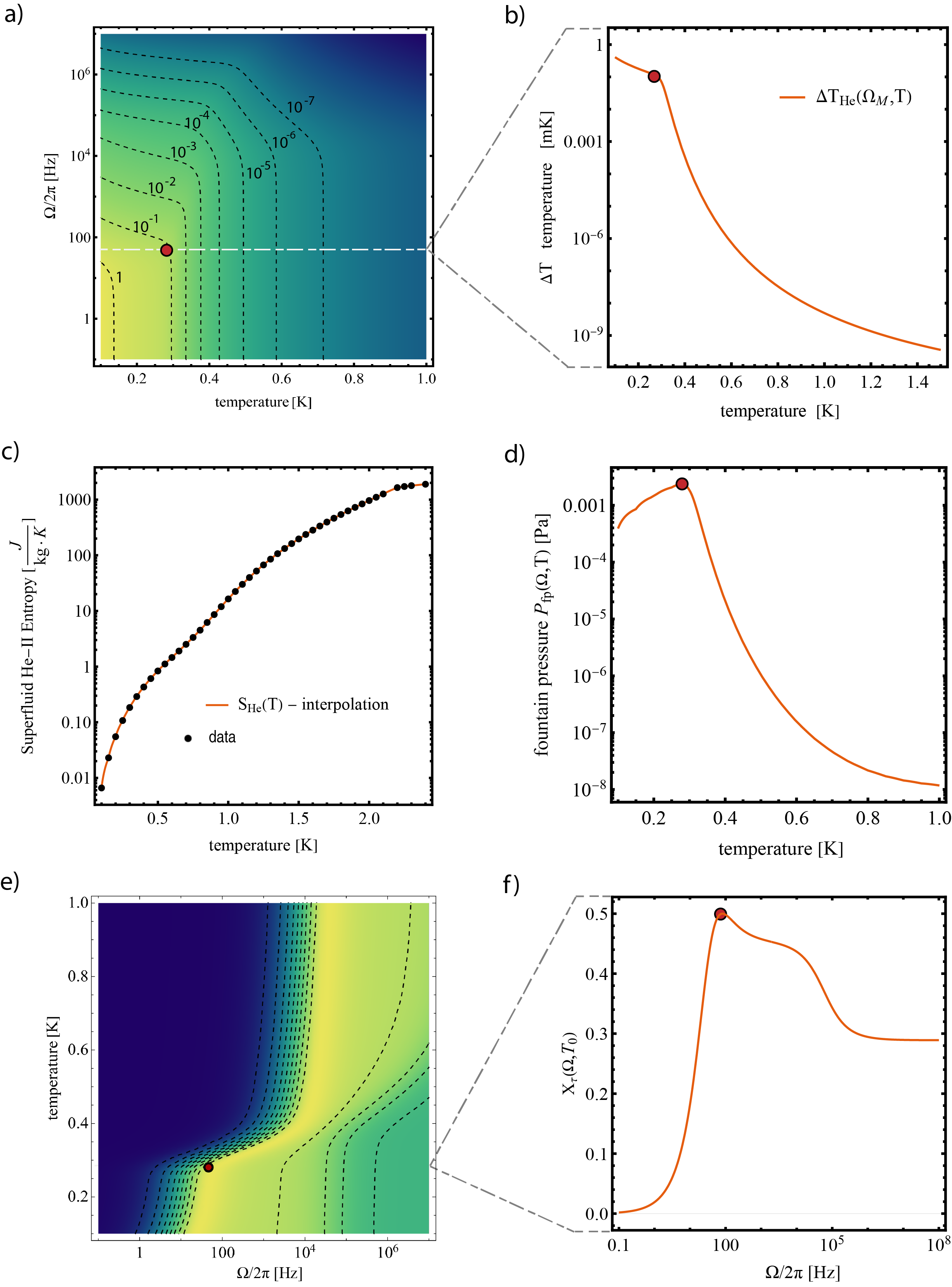}
    \caption{\textbf{Modelled transfer functions of the experimental system} In all figures the red dot represents the operation point of the experimental system. Figure a) shows a contour-color plot of the superfluid helium temperature increase $\Delta T_{\mathrm{He}}(\Omega,T)$ and its dependency on the thermal bath temperature $T$ and the modulation frequency $\Omega$. b) is a line plot of $\Delta T_{\mathrm{He}}(\Omega_{\mathrm{M}},T)$ at the experimental mechanical frequency $\Omega_{\mathrm{M}}$. Figure c) shows the interpolation function $S_{\mathrm{He}}(T)$ along with the data of entropy of superfluid helium from \cite{donnelly_observed_1998}.  d) plots the superfluid helium fountain pressure $P_{\mathrm{fp}}(\Omega_M,T)$. e) is a contour-color plot of the dynamical backaction function $\chi_{\mathrm{He}}(\Omega,T)$ and f) is a 2D plot of the same function at the given bath temperature $T=284$\,mK.}
    \label{fig:transfer-functions}
\end{figure}
The combined impedance of superfluid helium and the Kapitza impedance is $\tilde{Z}_{\mathrm{KHe}}(\Omega,T)=\tilde{Z}_{\mathrm{He}}(\Omega,T) +Z_{\mathrm{K}}(T)$, as these two elements are in series, see Fig. \ref{fig:electric_circuit}. This leads to the total impedance of the circuit:
\begin{equation}
   \tilde{Z}_{\mathrm{tot}}(\Omega,T) =\left(\frac{1}{\tilde{Z}_{\mathrm{sub}}(\Omega,T)}+\frac{1}{\tilde{Z}_{\mathrm{KHe}}(\Omega,T)}\right)^{-1}.
   \label{Eq:imp_tot}
\end{equation}

Therefore the voltage change (thermal realm: temperature) in the substrate is given by Ohm's law:
\begin{equation}
   \tilde{V}_{\mathrm{sub}}(\Omega,T) =I_{\mathrm{ph}}\tilde{Z}_{\mathrm{tot}}(\Omega,T).
   \label{Eq:voltage_sub}
\end{equation}
Considering the temperature drop over the Kapitza resistance, the temperature difference in the superfluid helium can be written as:
\begin{equation}
\begin{split}
   \tilde{V}_{\mathrm{He}}(\Omega,T) & =\tilde{V}_{\mathrm{sub}}(\Omega,T)-V_{\mathrm{K}}(T) \\
   & = I_{\mathrm{ph}}\tilde{Z}_{\mathrm{tot}}(\Omega,T) - \tilde{I}_{\mathrm{KHe}}(\Omega,T)Z_{\mathrm{K}}(T).
   \end{split}
   \label{Eq:voltage_sf}
\end{equation}
The heat flow rate (electric-analogy: current) towards the superfluid helium  $\tilde{I}_{\mathrm{KHe}}(\Omega,T)$ is given by Kirchhoff's law as: 
\begin{equation}
   \tilde{I}_{\mathrm{KHe}}(\Omega,T) =I_{\mathrm{ph}}-\tilde{I}_{\mathrm{sub}}(\Omega,T),
   \label{Eq:current_KHe}
\end{equation}
Using equations (\ref{Eq:voltage_sub}), (\ref{Eq:current_KHe}) and the relation $\tilde{I}_{\mathrm{sub}}(\Omega,T)=\tilde{V}_{\mathrm{sub}}(\Omega,T)/\tilde{Z}_{\mathrm{sub}}(\Omega,T)$ in equation (\ref{Eq:voltage_sf}), results in:
\begin{equation}
\begin{split}
   \tilde{V}_{\mathrm{He}}(\Omega,T) & = I_{\mathrm{ph}}\tilde{Z}_{\mathrm{tot}}(\Omega,T) - \left(I_{\mathrm{ph}}-\frac{I_{\mathrm{ph}}\tilde{Z}_{\mathrm{tot}}(\Omega,T)}{\tilde{Z}_{\mathrm{sub}}(\Omega,T)}\right)Z_{\mathrm{K}}(T)\\
   & = I_{\mathrm{ph}}\left(\tilde{Z}_{\mathrm{tot}}(\Omega,T)-Z_{\mathrm{K}}(T)+\frac{Z_{\mathrm{K}}(T)\tilde{Z}_{\mathrm{tot}}(\Omega,T)}{\tilde{Z}_{\mathrm{sub}}(\Omega,T)}\right).
   \end{split}
   \label{Eq:Vsf2}
\end{equation}
The voltage here is a complex number. The real temperature increase in the superfluid helium film and the substrate are given by the modulus of the voltage:
\begin{equation}
    \Delta T_{\mathrm{He}} (\Omega,T)= V_{\mathrm{He}}(\Omega,T) = |\tilde{V}_{\mathrm{He}}(\Omega,T)|
   \label{Eq:THe}
\end{equation}
and
\begin{equation}
    \Delta T_{\mathrm{sub}} (\Omega,T)= V_{\mathrm{sub}}(\Omega,T) = |\tilde{V}_{\mathrm{sub}}(\Omega,T)|.
   \label{Eq:Tsub}
\end{equation}
Fig. \ref{fig:transfer-functions} (a) and (b) demonstrate $\Delta T_{\mathrm{He}} (\Omega,T)$ and its dependency on the bath temperature $T$ and the drive frequency $\Omega$ for an absorbed power of $I_{\mathrm{ph}}=1$\,pW, where the red dot displays the operation point of our experimental system. Together with the temperature dependent entropy $S_{\mathrm{He}}(T)$, which is given by an interpolation to some measured data from \cite{donnelly_observed_1998} (see figure \ref{fig:transfer-functions} c)), the fountain pressure takes the form:
\begin{equation}
    P_{\mathrm{fp}} (\Omega,T)=\rho_{\mathrm{He}}\, S_{\mathrm{He}}(T)\, \Delta T_{\mathrm{He}} (\Omega,T),
    \label{Eq:fountain_pressure}
\end{equation}
and is plotted in figure \ref{fig:transfer-functions} d) for the experimental mechanical frequency $\Omega_{\mathrm{M}}$. This model shows that our operation temperature of $T = 284$\,mK is the optimal temperature  that results in the maximum fountain pressure for our system. This maximum arises as a consequence of two competing trends. On one hand, the entropy is an increasing function of T, pointing towards a stronger fountain pressure interaction at higher temperatures (Fig.\ref{fig:transfer-functions}c). On the other hand, the increase in heat capacity and thermal conductivity at higher temperatures reduces the temperature rise $\Delta T$, counteracting the previous effect (Fig.\ref{fig:transfer-functions}b).
The fountain pressure force is finally:
\begin{equation}
    F_{\mathrm{fp}} (\Omega,T)=P_{\mathrm{fp}} (\Omega,T) \mathcal{A},
    \label{Eq:fountain_pressure_force}
\end{equation}
with $\mathcal{A}$ being the surface area of the whole resonator.

\subsection{Thermal response time}

In addition to maximising $\Delta T_{\mathrm{He}} (\Omega,T)$ to optimise the fountain pressure strength, it is also important to understand and optimize the dynamical backaction efficiency of the system. The theory of photothermal heating and cooling~\cite{metzger_self-induced_2008, restrepo_classical_2011} shows that strongest backaction is achieved in the regime $\Omega_{\mathrm{M}} \,\tau_{\mathrm{th}} \sim 1$, where $\tau_{\mathrm{th}}$ corresponds to the thermal response time of the superfluid film. The thermal time delay $\tau_{\mathrm{th}}$ is given by the relation $\tau_{\mathrm{th}} = \phi/\Omega$, with $\Omega$ the mechanical frequency and $\phi$ the phase  of the complex transfer function $\tilde{V}_{\mathrm{He}}(\Omega,T)$: 
\begin{equation}
    \phi _{\mathrm{He}}(\Omega,T)=\arg(\tilde{V}_{\mathrm{He}}(\Omega,T)),
    \label{Eq:phaseshift_circuit}
\end{equation}
which means that $\tau_{\mathrm{th}}$ is frequency- and temperature-dependent. The unitless functional form:
\begin{equation}
    \chi _{\mathrm{He}}(\Omega,T)=\frac{\Omega \tau_{\mathrm{th}}((\Omega,T))}{1+(\Omega \tau_{\mathrm{th}}((\Omega,T)))^2},
    \label{Eq:chi_omegatau}
\end{equation}
represents the optimal time delay of the bolometric forces~\cite{metzger_cavity_2004,metzger_optical_2008,restrepo_classical_2011,de_liberato_quantum_2011}. For $\Omega_{\mathrm{M}} \,\tau_{\mathrm{th}} \sim 1$ we get $\chi _{\mathrm{He}}(\Omega=1/\tau_{\mathrm{th}},T)=0.5$.  Fig. \ref{fig:transfer-functions}(e) shows $\chi _{\mathrm{He}}(\Omega,T)$ in a color-contour plot for our system. Fig. \ref{fig:transfer-functions}(f) is a line cut through (e)  for a fixed temperature of $284$~mK. The red dot marks the mechanical mode frequency $\Omega_{\mathrm{M}}$ of the system, which is at the maximum value of $0.5$ for $\chi _{\mathrm{He}}$. As a consequence, our choice of superfluid mechanical mode and cryostat temperature allows us to operate at both the optimal point for fountain pressure strength (Fig.\ref{fig:transfer-functions}d), and optimal time-delayed forcing for dynamical backaction (Fig.\ref{fig:transfer-functions}f).
\begin{figure}
    \centering
    \includegraphics[width=0.8\textwidth]{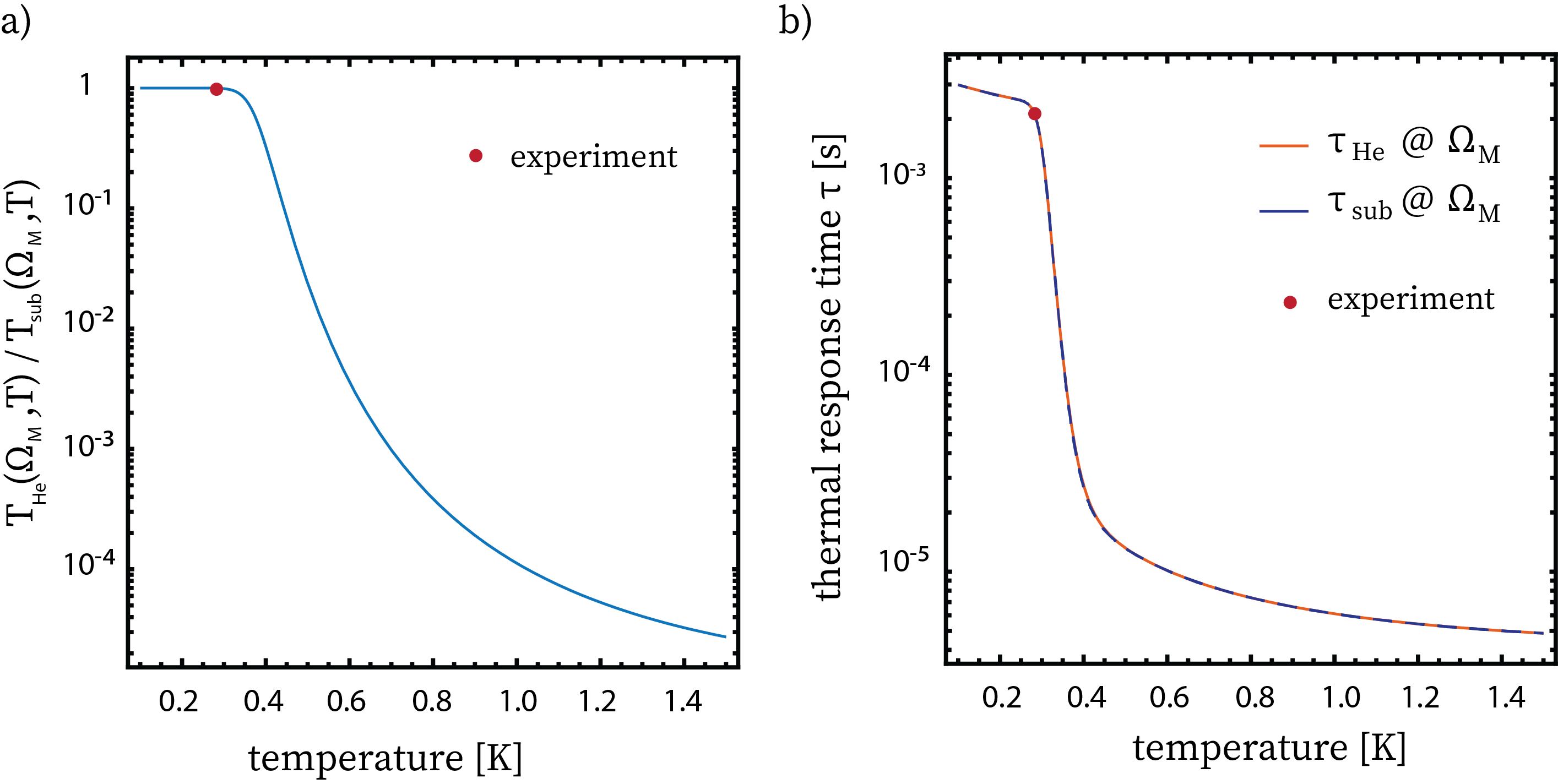}
    \caption{\textbf{Temperature ratio and thermal response time of superfluid helium and the substrate}. At the frequency and temperature used in the experiments (red dot), the temperature of the film closely tracks that of the underlying silica resonator, allowing these to be modelled as a common element in the ODE simulations of section \ref{suppsectionnumericalmodelODE}. }
    \label{fig:responsetime}
\end{figure}

\subsection{Figure of merit - fountain pressure dynamical backaction optimization}

The two main dynamical forces in our system are the fountain pressure force and the radiation pressure force given by:
\begin{equation}
    F_{\mathrm{rad}}=n_{\mathrm{cav}} \hbar G,
    \label{Eq:chi_omegatau_RadPressure}
\end{equation}
where $n_{\mathrm{cav}}$ is the intracavity photon number, $\hbar$ the reduced Planck constant and $G$ the optomechanical coupling rate. Having the two forces, fountain pressure force and radiation pressure force, and the dynamical backaction efficiency $\chi _{\mathrm{He}}(\Omega,T)$ leads to a figure of merit for the photothermal effect in our system:
\begin{equation}
    f_{\mathrm{M}}(\Omega,T)=\frac{F_{\mathrm{fp}}(\Omega,T)}{F_{\mathrm{rad}}} \, \chi _{\mathrm{He}}(\Omega,T).
    \label{Eq:fig_merit}
\end{equation}

\section{Numerical model}
\label{suppsectionnumericalmodelODE}

\subsection*{Differential equations}
The dynamical behaviour of the superfluid resonator may be described by three coupled differential equations relating to the intracavity photon number $n_{\mathrm{cav}}$, the change in mean film thickness at the level of the WGM $x$ and the  temperature $T$. Each of these parameters respectively responds on a characteristic timescale of $1/\kappa\sim$ ns; $\tauth\sim$ ms and $1/\Gamma\sim$ s. Since the optical decay rate $\kappa$ is much larger than all other decay rates, we consider that the intracavity photon number $n_{\mathrm{cav}}$ reacts instantaneously to any changes in the cavity (adiabatic limit), such that it takes the steady-state form~\cite{aspelmeyer_cavity_2014}: 
\begin{equation}
    \left| a\right|^2 =n_{\mathrm{cav}}=\frac{\kappaex}{\Delta^2+\left(\frac{\kappa}{2}\right)^2}\frac{P}{\hbar \omega_L},
    \label{Eq_ncav}
\end{equation}
where $\kappa=\kappaex+\kappa_i$ is the sum of the extrinsic and intrinsic loss rates respectively \cite{aspelmeyer_cavity_2014}, and $P$ the laser power at the level of the fiber taper. The detuning $\Delta$ is equal to:
\begin{equation}
    \Delta=\Delta_0+G x\quad \mathrm{with} \quad G=\frac{\partial \omega_0}{\partial x},
\end{equation}
with $\Delta_0$ the cavity detuning for zero displacement and   $G$ the optomechanical coupling rate (see section \ref{appendixsectionG}).
The dynamics can thus be reduced to two coupled equations of motion. The first determines the motion of the superfluid film:
 \begin{equation}
    \meff \,\ddot{x}+\meff \,\Gamma \,\dot{x}+\meff \,\Omega^2\, x=F_{\mathrm{fp}}=\rho\, S(T_0)\, (T-T_0)\, \mathcal{A},
    \label{Eq_eq_of_motion_mech}
\end{equation}
where $T_0$  and  $T$ are respectively the temperature of the environment and that of the superfluid film covering the resonator. This equation, which assumes a constant value for the entropy $S(T_0)$ is valid in the limit $\Delta T\ll T_0$, which is the case in the experiments.
The second governs the evolution of the temperature $T$ and arises from conservation of energy:
\begin{equation}
    \dot{T}=\frac{n_{\mathrm{cav}}\,\hbar\,\omega_L\,\kappa_i\,\alpha_{\mathrm{abs}}}{m\,c}-\frac{G_{\mathrm{th}}(T_0) \,(T-T_0)}{m\,c}
    \label{Eq_eq_of_motion_therm}
\end{equation}

Here, $\alpha_{\mathrm{abs}}\in [0,1]$ corresponds to the fraction of the intrinsic losses dissipated as heat in the resonator, $m$ to the resonator's thermal mass and $c$ its specific heat capacity, and $G_{\mathrm{th}}=\frac{m\,c}{\tau_{\mathrm{th}}}$ the resonator's thermal conductance.
We note here that at the operational point used in the experiments ($\Omega/2\pi=72$ Hz; $T=284$ mK) and the low optical powers in the pW range, the superfluid film temperature closely tracks that of the silica microsphere with minimal temperature difference and phase lag (verified through the thermal model of section \ref{suppsectionthermalequivalentcircuit}, and plotted in Fig.~\ref{fig:responsetime}). For this reason, in these time-domain numerical simulations, we simplify the thermal system by considering the silica resonator and superfluid film as a common element, of mass $m$ and heat capacity $c$, dominated by the microsphere mass and heat capacity. 
This allows us to accurately reproduce the experimental results, as shown below. 
The time dynamics of our system are obtained by numerically solving the coupled differential equations (Eqs \eqref{Eq_eq_of_motion_mech} and \eqref{Eq_eq_of_motion_therm}) with an ODE solver (MATLAB software).

\subsection*{Numerical simulations results}
\begin{figure}
    \centering
    \includegraphics[width=0.9\textwidth]{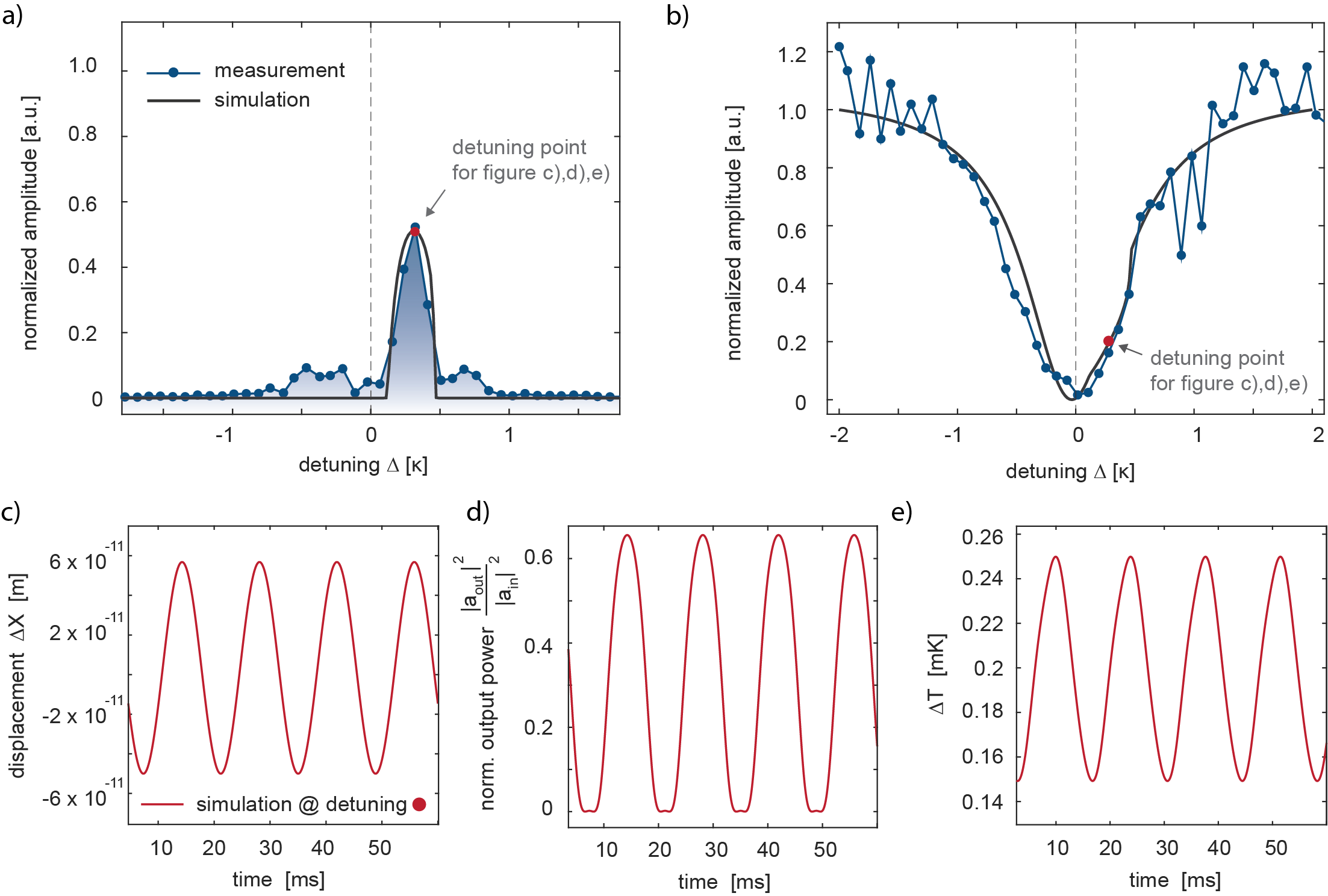}
    \caption{\textbf{Numerical simulation for 3.4\,pW input power.} a) Mechanical mode ($\Omega/2\pi=72$~Hz) amplitude as a function of cavity detuning, where the blue dots represent the normalized measured mode peak of the power spectrum and the black line the numerical simulations. b) Normalized optical transmission of the whispering gallery mode resonator depending on the detuning, obtained by plotting the experimentally measured normalized spectral density peak value of the calibration peak at 180\,Hz (blue dots) and the local oscillators (80\,MHz) normalized spectral density peak value in the numerical simulations (black line). The numerical simulations in c), d) and e) show the time dependency in the steady state regime of respectively the displacement $\Delta X$, the cavity transmitted normalized optical power and the temperature fluctuation $\Delta T$ around $T_{0}=284$\,mK. These simulations are performed at the detuning  represented by the red dot in  a) and b).}
    \label{fig:3p4pW-deltaX_deltaT}
\end{figure}

\begin{figure}
    \centering
    \includegraphics[width=0.9\textwidth]{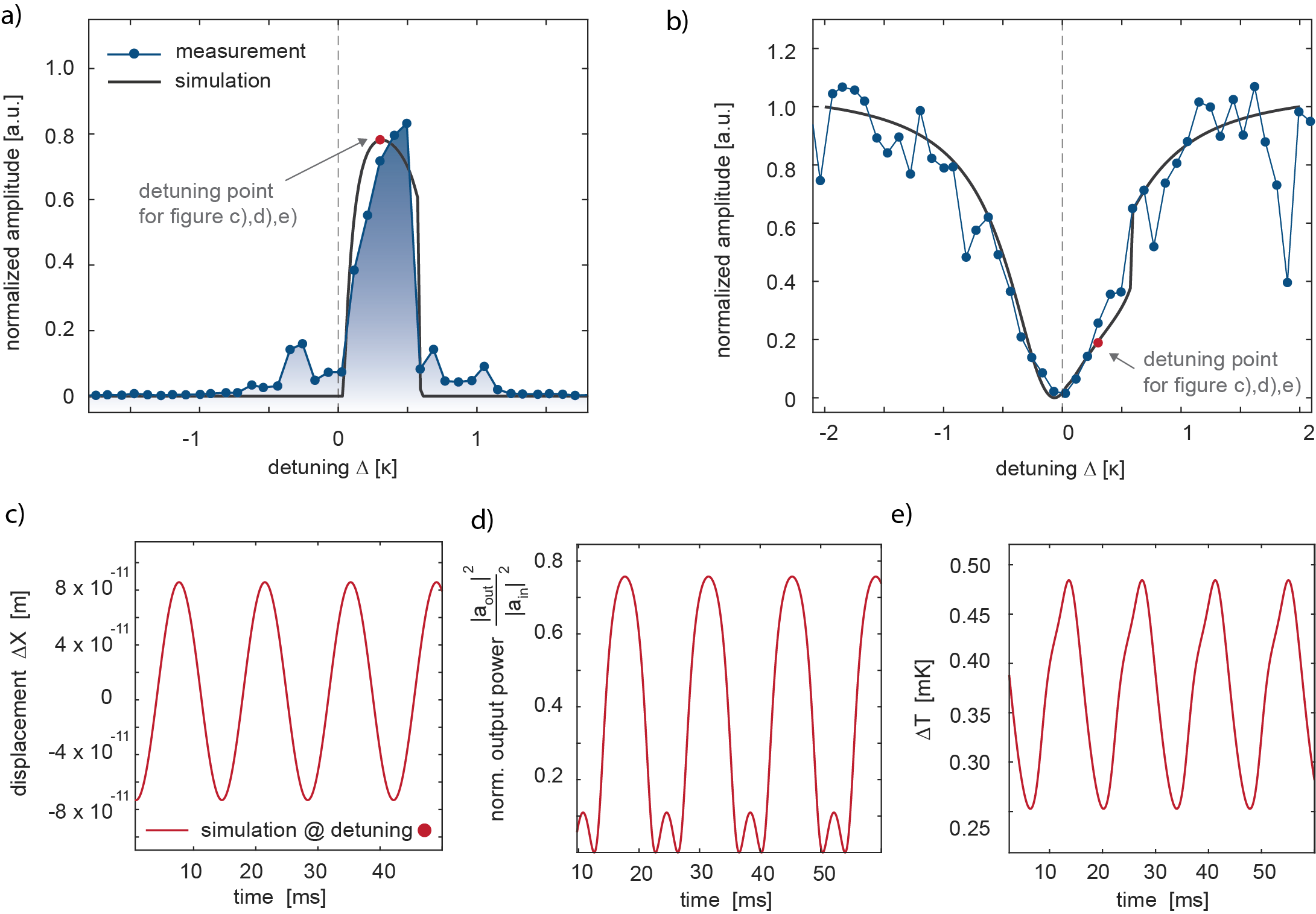}
    \caption{\textbf{Numerical simulation for 6.8\,pW input power.} a) Mechanical mode ($\Omega/2\pi=72$~Hz) amplitude as a function of cavity detuning, where the blue dots represent the normalized measured mode peak of the power spectrum and the black line the numerical simulations. b) Transmitted optical power from of the whispering gallery mode depending on the detuning, obtained by plotting the experimentally measured normalized spectral density peak value of the calibration peak at 180\,Hz (blue dots) and the local oscillators (80\,MHz) normalized spectral density peak value in the numerical simulations (black line). The numerical simulations in c), d) and e) show the time dependency in the steady-state regime  of respectively the displacement $\Delta X$, the cavity transmitted normalized optical power and the temperature fluctuation $\Delta T$ around $T_{0}=284$\,mK. These simulations are performed at the detuning  represented by the red dot in   a) and b).}
    \label{fig:6p8pW-deltaX_deltaT}
\end{figure}

\begin{figure}
    \centering
    \includegraphics[width=0.9\textwidth]{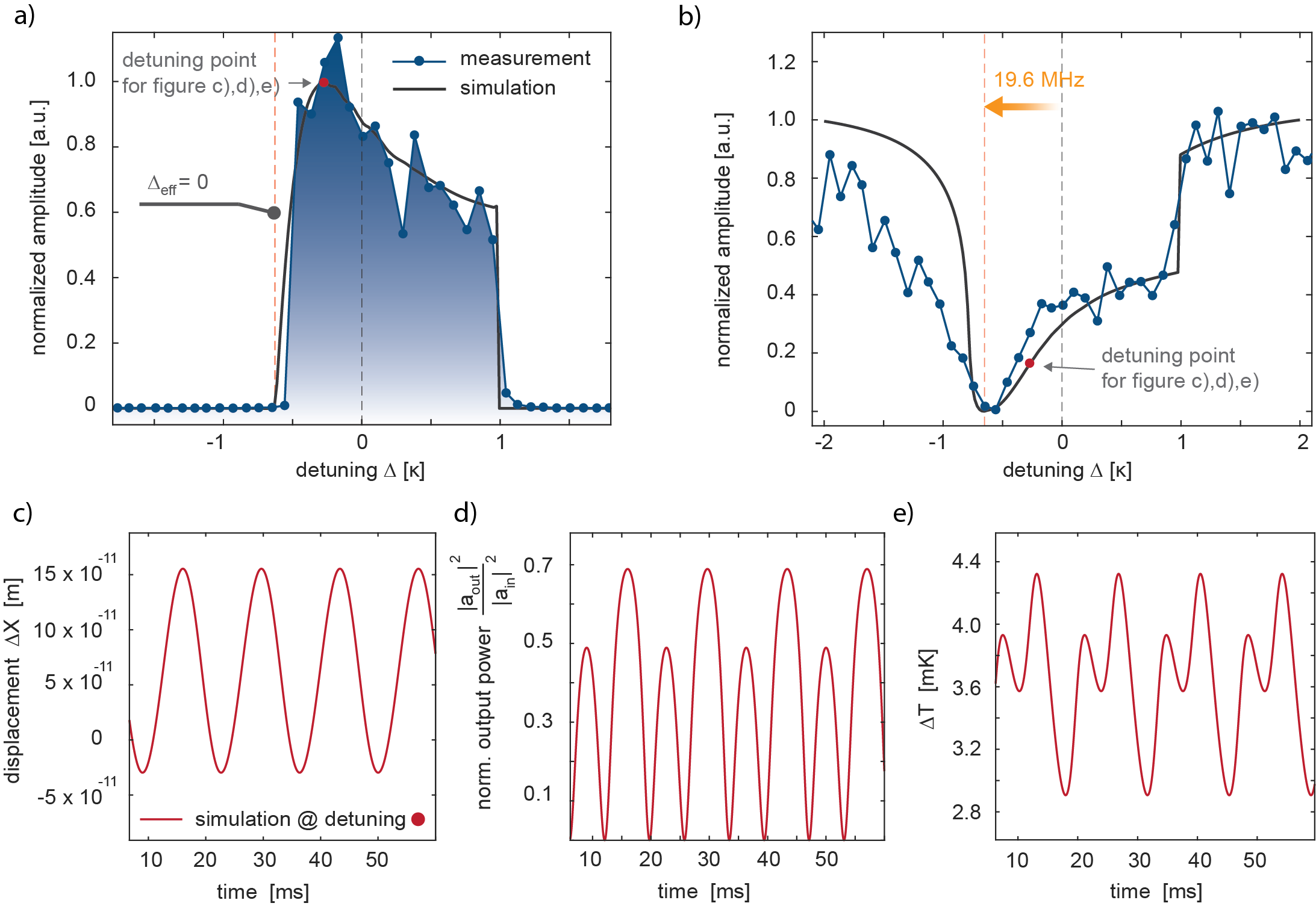}
    \caption{\textbf{Numerical simulation for 68\,pW input power.}a) shows the mechanical mode ($\Omega/2\pi=72$~Hz) amplitude depending on the cavity detuning, where the blue dots represent the normalized measured mode peak of the power spectrum and the black line the numerical simulations. b) Normalized optical transmission of the whispering gallery mode resonator as a function of detuning, obtained by plotting the experimentally measured normalized spectral density peak value of the calibration peak at 180\,Hz (blue dots) and the local oscillators (80\,MHz) normalized spectral density peak value in the numerical simulations (black line). At this high power, the D fountain pressure force causes the superfluid film to thicken, which leads to an optical resonance shift of 19.6\,MHz.  The numerical simulations in c), d) and e)  show the time dependency in the steady state regime of respectively the displacement $\Delta x$, the cavity transmitted normalized optical power and the temperature fluctuation $\Delta T$ around $T_{0}=284$\,mK.  These simulations are performed at the detuning  represented by the red dot in   a) and b).}
    \label{fig:68pW-deltaX_deltaT}
\end{figure}

Solving these equations with the parameters provided in Table \ref{Table_physical_params}, we obtain the dynamical behaviour of the superfluid film displacement $\Delta x(t)$, normalized optical output power $|a_{\mathrm{out}}(t)|^2/|a_{\mathrm{in}}|^2$ (where $|a_{\mathrm{out}}(t)|^2$ and $|a_{\mathrm{in}}(t)|^2$ are respectively the output and input optical powers, and are related via input-output formalism~\cite{haus1984waves,aspelmeyer_cavity_2014}) and the temperature fluctuations $\Delta T(t)$ around the environment temperature of $T=284$\,mK. The sub-figures c), d) and e) in figures \ref{fig:3p4pW-deltaX_deltaT}, \ref{fig:6p8pW-deltaX_deltaT} and \ref{fig:68pW-deltaX_deltaT} show the time-dependency of these parameters in the steady-state phonon lasing regime (i.e. after the initial transient dynamics) at the input powers 3.4\,pW, 6.8\,pW and 68\,pW.  The simulations are performed at a cavity detuning where the superfluid film's motional amplitude is maximized (represented by the red dot in sub-figure a)).
To appropriately compare the numerical simulations with the measurements, done with a spectrum analyzer and with a heterodyne detection scheme, we added an optical local oscillator field and applied a Fourier transformation to the time-dependent normalized optical output field in the steady state regime of the system for different cavity detunings and optical input powers. The black line in all sub-figures a) of figures \ref{fig:3p4pW-deltaX_deltaT}, \ref{fig:6p8pW-deltaX_deltaT} and \ref{fig:68pW-deltaX_deltaT} demonstrates the peak value of the fundamental mechanical mode in the normalized spectral density versus cavity detuning for different input powers. This we compared to the normalized amplitudes of the measured mechanical mode at different cavity detunings as well and same input powers, which are displayed as blue dots in all sub-figures a) of  figures \ref{fig:3p4pW-deltaX_deltaT}, \ref{fig:6p8pW-deltaX_deltaT} and \ref{fig:68pW-deltaX_deltaT}. The numerical simulation is fitted to the experimental data by using $\alpha_{\mathrm{abs}}$, the fraction of dissipated optical power that gets converted into heat in the resonator, and the mechanical mode decay rate $\Gamma$ as fitting parameters. The best results could be obtained with $\alpha_{\mathrm{abs}}=0.35$ and  $\Gamma$ being 1\,Hz, 1.5\,Hz and 5\,Hz respectively for the input powers  3.4\,pW, 6.8\,pW and 68\,pW, pointing towards an increase in the intrinsic damping rate with laser power. Such nonlinear damping has already been reported in the context of superfluid helium films~\cite{browne1984nonlinear}, as well as silicon optomechanical crystals~\cite{meenehan_silicon_2014}. In addition to the strong dynamical back action, induced by the photo thermal effect, we observe a static photothermal effect in the simulations as well. As explained in the main text, this static effect is caused by the rise of the mean temperature of the superfluid film. In sub-figure e) of the figures \ref{fig:3p4pW-deltaX_deltaT}, \ref{fig:6p8pW-deltaX_deltaT} and \ref{fig:68pW-deltaX_deltaT} it is shown that the mean temperature rises by about 0.2\,mK, 0.37\,mK and 3.6\,mK respectively for the input powers 3.4\,pW, 6.8\,pW and 68\,pW. This increase in temperature causes the superfluid film to thicken,  shifting the optical mode. For the input power of 68\,pW the optical mode is shifted by 19.6\,MHz, corresponding to an optical tunability of 288\,GHz/$\mathrm{\mu}$W, as shown in Fig.~\ref{fig:68pW-deltaX_deltaT}(b). This DC thicknening of the film is also apparent in Fig.~\ref{fig:68pW-deltaX_deltaT}(c), which shows the superfluid film oscillations around a new, thicker equilibrium position.
\begin{table}
    \centering
    \begin{tabular}{p{5 cm} c c c c }
    \hline
      Parameter   &  Symbol & Value & Unit & Source\\
      \hline
      WGM intrinsic energy decay rate & $\kappa_i/2\pi$ & 15 & MHz & measurement \\
      WGM extrinsic energy decay rate & $\kappaex/2\pi$ & 15 & MHz & measurement \\
      \hline
      Third sound mode frequency & $\Omega_M/2\pi$ & $72$ & Hz & measurement\\
      & & 86 & Hz & FEM\\
      Third sound mode effective mass & $\meff$ & $5.1\times 10^{-3}$& kg & FEM\\
      Third sound mode decay rate & $\Gamma$ & $2\pi\times 1$ & Hz & measurement\\
      Optomechanical coupling strength & $G/2\pi$ & $0.2\pm 0.01$ & GHz/nm & FEM\\
      Single photon optomechanical coupling rate & $g_0/2\pi$ & 0.7 & Hz & FEM\\
      Microsphere radius & $R$ & 49.5 &$\mu$m & SEM\\
      Mean superfluid film thickness & $d_0$ & 24 & nm & measurement\\ 
      Superfluid $^4$He density   & $\rho_{\mathrm{He}}$ & 145 & kg/m$^3$ & \cite{donnelly_observed_1998}\\
      First sound speed in superfluid helium & $c_{1_{\mathrm{He}}}$ & 236 & m/s & \cite{Atkins1951} \\
     superfluid helium mass (covering the total microsphere+stem) & $m_{\mathrm{He}}$ &  $2.65\times 10^{-12}$ & kg & FEM\\
      Silica density & $\rho_{\mathrm{SiO_2}}$ & 2200 & kg/m$^3$ & \cite{noauthor_comsol_nodate} \\
      Silica thermal conductivity (@284 mK) & $\kappa_{\mathrm{SiO_2}}$ &  $1.6 \times 10^{-3}$  & W/m/K & \cite{raychaudhuri_origin_1989} \\
      Silica specific heat capacity (@284 mK) &  $c_{\mathrm{SiO_2}}$ & $3.4 \times 10^{-4}$   & J kg$^{-1}$ K$^{-1}$ & \cite{enss_low-temperature_2005} \\
      Thermal conductance (@284mK) & $G_{\mathrm{th-sub}}$ & $4.1\times 10^{-9}$& W/K & FEM\\
      Silica microsphere area (incl. stem) & $\mathcal{A}$ & $7.7\times10^{-7}$ & m$^2$ & SEM\\
      Silica microsphere mass (incl. stem) & $m_{\mathrm{sub}}$ &  $4.9\times 10^{-8}$ & kg & FEM\\
      Longitudinal sound velocity of silica & $c_{l_{\mathrm{SiO_2}}}$ & 5968 & m/s & \cite{rumble2017crc} \\
      Transverse sound velocity of silica & $c_{t_{\mathrm{SiO_2}}}$ & 3764 & m/s & \cite{rumble2017crc} \\
      Thermal response time & $\tau_{\mathrm{th}}$ & 5.7 & ms & FEM\\
      He II entropy per unit mass (@284 mK) & S & 0.16  & J/kg/K & \cite{donnelly_observed_1998}\\
      fraction of $\kappa_i$ dissipated as heat & $\alpha_{\mathrm{abs}}$ & 0.35 &- & fit\\
      Operating temperature & $T$ & 284 & mK & measurement\\
         \hline
    \end{tabular}
    \caption{Physical parameters used in the simulation. FEM:Finite Element modelling. SEM: Scanning electron microscope. }
    \label{Table_physical_params}
\end{table}

\clearpage

\section{Lasing thresholds for various systems}
\label{comparelasingthreshold}
We benchmark our phonon lasing threshold against existing literature in Fig. 5 of the main text. We condensed the amount of systems shown in Fig. 5 to 14 different experiments, which are representative for the majority of phonon lasing systems and driving mechanisms (radiation pressure, electrostriction, photothermal and electrothermal interactions). All literature references that have been used for Fig. 5 are listed in Table \ref{table:comparinglasingsystems}.
\begin{sidewaystable}
    \centering
    \begin{tabular}{c r r r c l}
    \hline
      \#  &  \multicolumn{1}{c}{mechanical}  & \multicolumn{1}{c}{effective} & \multicolumn{1}{c}{lasing} &  \multicolumn{1}{c}{driving} & \multicolumn{1}{c}{Source}\\
       &  \multicolumn{1}{c}{frequency} & \multicolumn{1}{c}{mass} & \multicolumn{1}{c}{threshold} &  \multicolumn{1}{c}{mechanism} & \\
      \hline
      1 & 90\,MHz & $2.7\times 10^{-21}$\,kg & $\sim$ 5\,pW & electrothermal & C. Urgell, et al, \textit{Nature Physics} vol.\textbf{16}, pp.32–37 (2020)\\
      2 & 230\,MHz & $2.2\times 10^{-20}$\,kg & $\sim$ 10\,pW & electrothermal & Y. Wen, et al,\textit{Nature Physics} vol.\textbf{16}, pp.75–82 (2020)\\
      3 & 3.5\,MHz & $2\times 10^{-16}$\,kg & 500\,nW & photothermal & R.A. Barton, et al, \textit{Nano Letters} vol.\textbf{12}, \textbf{9}, pp. 4681–4686 (2012)\\
      4 & 3\,MHz & $2.1\times 10^{-16}$\,kg & 660\,nW & photothermal & R.De Alba, et al, \textit{Nano Letters} vol.\textbf{17}, \textbf{7}, pp.3995–4002 (2017)\\
      5 & 180\,kHz & $1.6\times 10^{-12}$\,kg & 30\,$\mu$W & photothermal & D. Woolf, et al, \textit{Optics Express} vol.\textbf{21}, \textbf{6}, pp. 7258-7275 (2013)\\
      6 & 8\,kHz & $2.3\times 10^{-12}$\,kg & 26\,$\mu$W & photothermal & C. Metzger, et al, \textit{Physical Review Letters} vol.\textbf{101}, 133903 (2008)\\
      7 & 7.3\,MHz & $1.2\times 10^{-14}$\,kg & 1.8\,$\mu$W & photothermal/radiation pressure & X. He, et al, \textit{Nature Physics} vol.\textbf{16}, pp.417–421 (2020)\\
      8 & 314\,MHz & $5.3\times 10^{-14}$\,kg & $\sim$50\,$\mu$W & photothermal/radiation pressure & P.E. Allain, et all, \textit{Physical Review Letters} vol.\textbf{126}, 243901 (2021)\\
      9 & 3\,GHz & $2.7\times 10^{-16}$\,kg & 2.5\,mW & radiation pressure & L. Mercade, et al, \textit{Nanophotonics} vol.\textbf{9}(11), pp.3535–3544 (2020)\\
      10 & 3\,GHz & $1.0\times 10^{-15}$\,kg & 40\,$\mu$W & photoelastic & I. Ghorbel, et al, \textit{APL Photonics} vol.\textbf{4}, pp.116103 (2019)\\
      11 & 2\,MHz & $7.6\times 10^{-15}$\,kg & 60\,$\mu$W & radiation pressure & H. Jayakumar, et al, \textit{Phys. Rev. Applied} vol.\textbf{16}, 014063 (2021) \\
      12 & 23\,MHz & $5.0\times 10^{-11}$\,kg & 7\,$\mu$W & radiation pressure & I. Grudinin, et al, \textit{Physical Review Letters} vol.\textbf{104}, 083901 (2010)\\
      13 & 1.2\,GHz & $5.7\times 10^{-15}$\,kg & 3\,$\mu$W & radiation pressure & W.C. Jiang, et al, \textit{Optics Express} vol.\textbf{20},14, pp. 15991-15996 (2012)\\
      14 & 8\,MHz & $1.5\times 10^{-13}$\,kg & 270\,nW & radiation pressure & Q. Lin, et al, \textit{Physical Review Letters} vol.\textbf{103},  103601 (2009)\\
   15 & 3.7 GHZ & $3.1\times 10^{-16}$\,kg & $\sim500$\,nW & radiation pressure/electrostriction & A. Krause et al, \textit{Physical Review Letters} vol.\textbf{115},  233601 (2015)\\
       16 & 3.6 GHz & $\sim0.3\times 10^{-15}$\,kg & 7\,nW & radiation pressure/electrostriction & S. Meenehan et al, \textit{Phys. Rev. A} vol.\textbf{90},  011803 (2014)\\
       \hline
      \hline
    \end{tabular}
    \caption{Sources used to compare the lasing thresholds for various systems in Fig. 5 of the main text.}
    \label{table:comparinglasingsystems}
\end{sidewaystable}

\section{Thermodynamic efficiency}
\label{sectionsuppthermodynamicefficiency}
The thermodynamic efficiency is estimated by multiplying the stored mechanical power with the mechanical damping rate, yielding an acoustic power loss $P_{\mathrm{mech}}$, which must be exactly compensated by the optical drive to maintain constant amplitude self-sustained oscillations. 
\begin{equation}
    P_{\mathrm{mech}}=\frac{1}{2}\meff \Omega^2 x^2 \Gamma
\end{equation}
Using the values from Table~\ref{Table_physical_params} and a displacement amplitude of $x=6\times 10^{-11}$ (obtained from Fig.~\ref{fig:3p4pW-deltaX_deltaT}), we get $P_{\mathrm{mech}} = 1.2\times 10^{-17}$\,W; while the dissipated optical power is given by $P_{\mathrm{abs}}= P \times \alphaabs = 3.4$\,pW $\times 0.35 =1.2$\,pW. This results in a thermodynamic efficiency (rate of conversion of heat into mechanical work) of $\eta=P_{\mathrm{mech}}/P_{\mathrm{abs}}=1\times 10^{-5}$. 

As a comparison we calculate the thermodynamic efficiency for the carbon nanotube electrothermal system with the closest lasing threshold ($\sim 5$\,pW)~\cite{urgell_cooling_2020}. The mechanical power produced to sustain phonon-lasing is:
\begin{equation}
    P^{\mathrm{nano}}_{\mathrm{mech}}=\frac{1}{2}\meff^{\mathrm{nano}} \Omega_{\mathrm{nano}}^2 x_{\mathrm{nano}}^2 \Gamma_{\mathrm{nano}}=4.3 \times 10^{-19}\,\mathrm{W},
\end{equation}
with the effective mass $\meff^{\mathrm{nano}}=2.7\times 10^{-21}$\,kg, the mechanical frequency  $\Omega_{\mathrm{nano}}/(2\pi)=90$\,MHz, the displacement $x_{\mathrm{nano}}=4$\,nm and the mechanical decay rate $\Gamma_{\mathrm{nano}}/(2\pi)=10$\,Hz~\cite{urgell_cooling_2020}. The electrical power dissipated as heat required for the production of $P^{\mathrm{nano}}_{\mathrm{mech}}$  is approximately $P_{\mathrm{elec}}=5$\,pW. This results in an thermodynamic efficiency for the nanotubes $\eta_{\mathrm{nano}}=P^{\mathrm{nano}}_{\mathrm{mech}}/P_{\mathrm{elec}}=0.86\times 10^{-7}$.
This comparison indicates our system is $\sim 10^{2}$ times more thermodynamically efficient than the nanotube system with the lowest reported lasing threshold, aside from ours.

\section{Single photon detection}
\label{sectionsuppsinglephotondetection}
The fountain pressure between a region of the film at temperature T, and a bath at temperature $T_0$ corresponds to
\begin{equation}
    P_{\mathrm{fp}}=\rho \, \int_{T_0}^T S(T') \, \mathrm{d}T'
\end{equation}
For a small difference in temperature, the fountain pressure is given by \cite{london_thermodynamics_1939, enss_low-temperature_2005}:
\begin{equation}
    P_{\mathrm{fp}}=\rho S\,\Delta T
    \label{Eqfountainpressure}
\end{equation}
At low temperatures, the specific heat of superfluid helium is several orders of magnitude larger than that of the underlying optical resonator material (e.g. silicon, silica)~\cite{pobell_matter_2007}. For instance, at 250 mK, $c_{\mathrm{He}}\simeq 1000\,c_{\mathrm{SiO2}}$. Therefore for miniature optical resonators, where the thickness of the superfluid film is no longer negligible compared to that of the resonator (consider e.g. a 30 nm thick superfluid film on either side of a 200 nm thick resonator), the entire thermal mass of the superfluid-covered resonator is dominated by the superfluid film. The temperature increase for a deposited energy $Q$ corresponding to one absorbed photon is therefore given by:
\begin{equation}
    \Delta T=\frac{Q}{m \,c_{\mathrm{He}}}=\frac{\hbar \omega}{m\, c_{\mathrm{He}}},
    \label{EqDeltaT}
\end{equation}
where m is the mass of the superfluid film, and $c_{\mathrm{He}}$ superfluid helium's specific heat capacity. 
Since the superfluid helium entropy takes the form~\cite{donnelly_observed_1998}:
\begin{equation}
        S=\int \frac{c_{\mathrm{He}}}{T} dT,
\end{equation}
for $c_{\mathrm{He}}$ of the form $\alpha T^3$, 
\begin{equation}
    S=\int\frac{\alpha T^3}{T} dT=\int\alpha T^2 dT=\frac{\alpha T^3}{3} \simeq\frac{c_{\mathrm{He}}}{3}
\end{equation}
(Which is indeed what is measured in practice, see ref. \cite{donnelly_observed_1998}).
Using this, we can re-express Eqs. (\ref{Eqfountainpressure}) and (\ref{EqDeltaT}) as:
\begin{equation}
    P_{fp}=\frac{\rho S(T) \hbar \omega}{\rho V c_{\mathrm{He}}(T)}
\end{equation}
and obtain a simple, temperature-independent expression for the induced fountain pressure resulting from the absorption of a photon of energy $\hbar \omega$:
\begin{equation}
\boxed{
    P_{\mathrm{fp}}\simeq\frac{\hbar \omega}{3\,V_{\mathrm{He}}}}
    \label{EqPfountainpressuresinglephoton}
\end{equation}
The goal is therefore to minimize the volume of superfluid covering the resonator, which is attained by reducing the resonator surface area. The above calculation is valid in the regime where:
\begin{itemize}
    \item The thermal response time of the resonator is slower than that of the superfluid film (which is on the order of a few microseconds at the temperatures considered here), allowing the thermal energy to be transferred to the superfluid film before it is lost to the environment through the resonator's own thermal anchoring.
    \item The thermal Kapitza conductance at the resonator/superfluid interface is well in excess of the thermal conductance between the helium film and its environment mediated through the vapor pressure. This allows heat to build up in the superfluid film before it is lost to the environment through evaporation. This criterion does not depend strongly on the choice of resonator material\footnote{Since the acoustic impedance mismatch between superfluid helium and most solids commonly  used for optical resonator fabrication (Silica, Silicon, Gallium Arsenide...) is of comparable magnitude.}, and is valid for temperatures below $\sim300$ mK. 
\end{itemize}

\begin{figure}
    \centering
    \includegraphics[width=0.7 \textwidth]{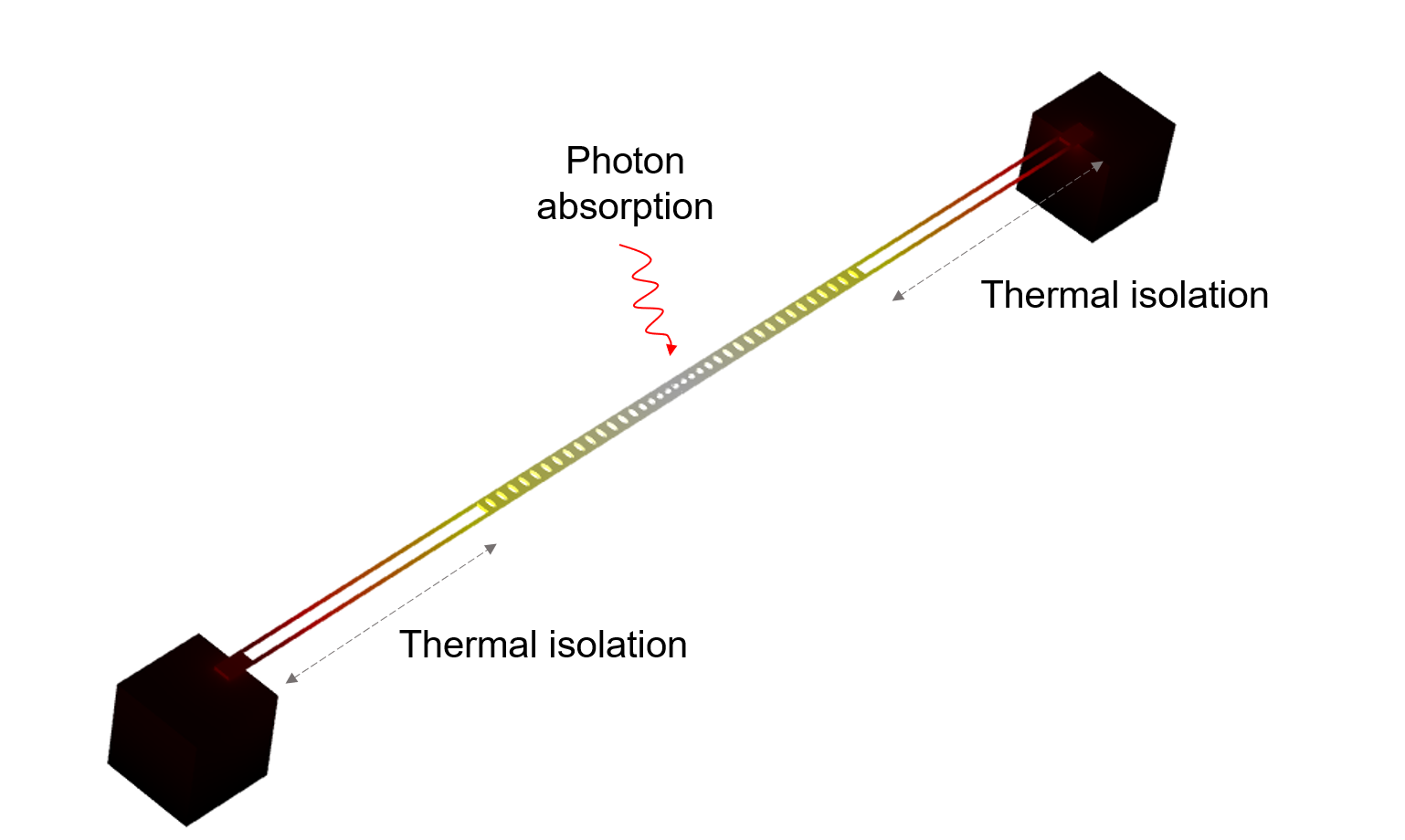}
    \caption{Silicon 1D photonic crystal architecture. Long tethers provide thermal isolation from the substrate. }
    \label{Figsinglephotondetection}
\end{figure}

An example architecture satisfying the above conditions is shown in Fig. \ref{Figsinglephotondetection}. It consists of a suspended 1D silicon optical crystal cavity, of the kind used in Ref.~\cite{chan_laser_2011}. Long and narrow tethers provide sufficient thermal isolation for the heat resulting from optical absorption to be predominantly communicated to the superfluid film.  
Such a resonator has a optomechanical coupling rate simulated through FEM of $G/2\pi=\sim5$ GHz/nm~\cite{baker_theoretical_2016}, (meaning that a
1 nm change in thickness of the superfluid film shifts the optical resonance frequency by 5 GHz). Using Eq. \ref{EqPfountainpressuresinglephoton}, we estimate a thickening of the film on the order of 1 nm is achievable per absorbed photon, leading to an optical shift well in excess of the optical resonator linewidth $\kappa$~\cite{chan_laser_2011}.

\subsection{Comparison to radiation pressure per photon} 

We compare the magnitude of the fountain pressure induced by a photon absorption event (Eq. \ref{EqPfountainpressuresinglephoton}) to the radiation pressure exerted by an intracavity photon. 
The radiation pressure due to a single photon acting on an element of the superfluid film ---which arises due to the change in electromagnetic energy density due to the presence of the helium---is given by:
\begin{equation}
    P_{\mathrm{rad}}=\frac{1}{2}\varepsilon_0\left(\epssf-1\right) E^2,
    \label{Eqradpressure}
\end{equation}
where the electric field is normalized such that:
\begin{equation}
    \frac{1}{2}\iiint \varepsilon_0\varepsilon_r(\vec{r}) E^2 \rmd V=\hbar \omega.
\end{equation}
Making the (strongly) simplifying assumption that the field be essentially localized within the optical resonator (of permittivity $\varepsilon_r$) and of constant magnitude over the mode volume $V_{\mathrm{mode}}$, leads to $E^2\sim \hbar\omega/(1/2\, \varepsilon_0\, \varepsilon_r\, V_{\mathrm{mode}})$. Combined with Eq. (\ref{Eqradpressure}), this provides an order of magnitude estimate of achievable the radiation pressure per photon acting on the superfluid film:
\begin{equation}
\boxed{
   P_{\mathrm{rp}}\simeq\frac{\varepsilon_{\mathrm{sf}}-1}{\varepsilon_r}\,\frac{\hbar \omega}{V_{\mathrm{mode}}}}
\end{equation}
Compared to Eq. (\ref{EqPfountainpressuresinglephoton}), we note the presence of the prefactor ($\frac{\varepsilon_{\mathrm{sf}}-1}{\varepsilon_r}=5\times 10^{-3}$ for a  silicon resonator), and the fact that the superfluid volume has been replaced by the (larger) optical mode volume. 
Combined, these two effects lead to an approximate 3 orders of magnitude reduction compared to the fountain pressure, see Eq. \eqref{EqPfountainpressuresinglephoton}.

\section{Comparison with thermo-elastic stress in a crystal}
\label{sectionsuppthermoelasticstress}
 The thermal stress $\sigma_{\mathrm{th}}$ arising from a temperature increase $\Delta T$ in an isotropic solid of bulk modulus K and thermal expansion coefficient $\alpha=\frac{1}{V}\left( \frac{\partial V}{\partial T}\right)$ is given by: 
\begin{equation}
    \alpha \Delta T=\frac{\sigma_{\mathrm{th}}}{K},
\end{equation}
where for one absorbed photon, $\Delta T$ is given by:
\begin{equation}
    \Delta T=\frac{\hbar \omega}{\rho V c}
\end{equation}
This yields a thermal stress per photon of
\begin{equation}
    \sigma_{\mathrm{th}}=\frac{\hbar\, \omega \,\alpha\, K}{\rho\, V\, c}
\end{equation}
This can be re-expressed in simpler form as:
\begin{equation}
    \sigma_{\mathrm{th}}=\gamma\frac{\hbar \omega }{V },
\end{equation}
where we have introduced the dimensionless Gr\"{u}neisen parameter $\gamma=\frac{\alpha K}{\rho c}$. For silicon at low temperatures, this takes a value of approximately $\gamma \simeq 0.2$~\cite{gauster_low-temperature_1971}, resulting in a thermal stress very close to the superfluid fountain pressure (Eq. (\ref{EqPfountainpressuresinglephoton})). It is thus primarily the compliance of the fluid interface, combined with the engineered ability to collect the thermal energy and operate near the optimal regime $\Omega \tau\sim 1$ which is responsible for the ultralow threshold observed here, and not a fundamentally stronger nature of the fountain pressure force in superfluid helium, making these results broadly applicable to photothermally- and electrothermally-driven systems.

\bibliography{Bibliography}
\bibliographystyle{ieeetr}